\documentclass[12pt,preprint]{aastex}

\usepackage{epsfig} 

\newcommand\etal{{et~al.}} 

\slugcomment{Accepted for publication in ApJ}
\shorttitle{On the metallicity discrepancies between GCs and their parent galaxies}
\shortauthors{Yoon et al.}

\begin{document}

\title{Nonlinear Color-Metallicity Relations of Globular Clusters. III. On the Discrepancy in Metallicity between Globular Cluster Systems and their Parent Elliptical Galaxies}

\author{Suk-Jin Yoon\altaffilmark{1}, Sang-Yoon Lee\altaffilmark{1}, John P. Blakeslee\altaffilmark{2}, Eric W. Peng\altaffilmark{3}, Sangmo T. Sohn\altaffilmark{4}, Jaeil Cho\altaffilmark{1}, Hak-Sub Kim\altaffilmark{1}, Chul Chung\altaffilmark{1}, Sooyoung Kim\altaffilmark{1}, and Young-Wook Lee\altaffilmark{1}}

\altaffiltext{1}{Department of Astronomy and Center for Galaxy Evolution Research, Yonsei University, Seoul 120-749, Korea}
\email{sjyoon@galaxy.yonsei.ac.kr}
\altaffiltext{2}{Herzberg Institute of Astrophysics, National Research Council of Canada, Victoria, BC V9E 2E7, Canada} 
\altaffiltext{3}{Department of Astronomy and Kavli Institute for Astronomy and Astrophysics, Peking University, Beijing 100871, China}
\altaffiltext{4}{Space Telescope Science Institute (STScI), 3700 San Martin Drive, Baltimore, MD 21218}

\begin{abstract}
One of the conundrums in extragalactic astronomy is the discrepancy in observed metallicity distribution functions (MDFs) between the two prime stellar components of early-type galaxies---globular clusters (GCs) and halo field stars. This is generally taken as evidence of highly decoupled evolutionary histories between GC systems and their parent galaxies. Here we show, however, that new developments in linking the observed GC colors to their intrinsic metallicities suggest nonlinear color-to-metallicity conversions, which translate observed color distributions into strongly-peaked, unimodal MDFs with broad metal-poor tails.
Remarkably, the inferred GC MDFs are similar to the MDFs of resolved field stars in nearby elliptical galaxies and those produced by chemical evolution models of galaxies. 
The GC MDF shape, characterized by a sharp peak with a metal-poor tail, indicates a virtually continuous chemical enrichment with a relatively short timescale. 
The characteristic shape emerges across three orders of magnitude in the host galaxy mass, suggesting a universal process of chemical enrichment among various GC systems.
Given that GCs are bluer than field stars within the same galaxy, it is plausible that the chemical enrichment processes of GCs ceased somewhat earlier than that of field stellar population, and if so, GCs preferentially trace the major, vigorous mode of star formation events in galactic formation. We further suggest a possible systematic age difference among GC systems, in that the GC systems in more luminous galaxies are older. This is consistent with the downsizing paradigm whereby stars of brighter galaxies, on average, formed earlier than those of dimmer galaxies; this additionally supports the similar nature shared by GCs and field stars. 
Although the sample used in this study (the {\it HST} ACS/WFC, WFPC2, and WFC3 photometry for the GC systems in the Virgo galaxy cluster)  confines our discussion to $R$ $\lesssim$ $R_e$ for giant ellipticals and $\lesssim$ 10 $R_e$ for normal ellipticals, our findings suggest that GC systems and their parent galaxies have shared a more common origin than previously thought, and hence greatly simplify theories of galaxy formation. 
\end{abstract}

\section{INTRODUCTION}

Globular cluster (GC) systems are always present in large galaxies. It is generally believed that GCs form when starbursts occur in galaxies, and as `fossil records', contain vital information on the formation and evolution of their parent galaxies \citep{searle78,harris91,ashman98,west04,brodie06,mglee10a}. 
Despite their close interplay between GCs and field stars, the comparative studies have uncovered a fundamental difference in the observed shapes of their metallicity distribution functions (MDFs), even within the relatively simple type of galaxies -- elliptical galaxies \citep{harris02,rejkuba05,harris07a,harris07b,bird10}.
The cause of the discrepancy between these two prime stellar components of galaxies has been the topic of much interest both on theoretical \cite[e.g.,][]{beasley02,pipino07} and observational grounds \cite[e.g.,][]{forbes01,forte05,forte07,liu11} because the disagreement signifies highly decoupled evolutionary paths between GC systems and their parent galaxies.

The most common technique for measuring metallicities of a substantially large sample of GCs is {\it photometry} -- obtaining their broadband colors, although it is no substitute for spectroscopy. Because GCs in the Milky Way and other galaxies are usually older than 10 Gyr and age does not strongly affect GC broadband colors of GCs this old, the main parameter governing GC colors is metallicity. Indeed, the overall, first-order feature of the color-metallicity relations (hereafter ``CMRs'') is that GC colors scale {\it linearly} with their metallicities. Empirical relationships between the most often used color, $V-I$, and [Fe/H], fitted mainly to the Galactic GCs \citep{couture90,kissler-patig97,kissler-patig98,kundu98,barmby00}, are approximately linear. Using the more metallicity-sensitive colors $C-T_1$ or $C-R$, both \citet{harris02} and \citet{cohen03} found a mildly quadratic or broken linear relationship between [Fe/H] and color to be a better fit, thus improving on the relation of \citet{geisler90}.

With the linear or mildly curved color-to-metallicity conversion, the now well-documented observation of bimodality in GC color distributions \cite[e.g.,][]{zepf93,ostrov93,whitmore95,mglee98,gebhardt99,harris01,kundu01,larsen01,peng06,harris06,jordan09,sinnott10,liu11} has been translated into bimodality of their MDFs. This is where a sharp distinction between GCs and field stars takes place; independent studies via {\it direct} photometry of spatially-resolved constituent field stars in a dozen nearby galaxies have shown that their MDFs have, in general, strongly-peaked, {\it unimodal} [Fe/H] distributions with broad metal-poor tails.

More recent observations and modeling of old star clusters, however, suggest that the relations between metallicity and broadband color for GCs have a subtle, second-order feature; they appear to be nonlinear with a quasi-inflection at intermediate metallicities. For instance, \citet{peng06} presented an empirical relationship between the $g-z$ colors and spectroscopic metallicities for GCs in the Milky Way and the giant elliptical galaxies, M49 and M87 (see their Figure 11). With this dataset, they showed that the relationship between [Fe/H] and $g-z$ is steep for [Fe/H] $<$ $-$0.8, shallow up to [Fe/H] $\simeq$ $-$0.5, and then possibly steep again at higher metallicities.

Independently, \citet[][hereafter Paper I]{yoon06} presented a theoretical metallicity-to-color relationship that has a significant inflection and thus reproduces well the observed feature. This nonlinear nature of the relation between intrinsic metallicity and its proxy, colors, may hold the key to understanding the color bimodality phenomenon. Paper I showed that the wavy feature projects equidistant metallicity intervals near the quasi-inflection point onto larger color intervals, and thus produces bimodal GC color distributions when the underlying distribution in [Fe/H] is broad, even if it is unimodal. The scenario gives a simple and cohesive explanation for the key observations, including ($a$) the overall shape of color histograms, ($b$) the number ratio of blue and red GCs as a function of host galaxy luminosity, and ($c$) the peak colors of {\it both} blue and red GCs as a function of host luminosity.
If the bona-fide shape of the color-metallicity relationship is highly inflected, what has been thought to be the MDFs of GC systems may deviate significantly from the true distributions.

In this paper, we present an alternative way of resolving the long-standing discrepancy in the MDFs between GCs and halo stars in bright elliptical galaxies. 
The nonlinear conversion from metallicities to colors (Paper I) should not be irreversible,
and here we try to {\it inverse-transform} color distributions of GCs into metallicity distributions 
using the nonlinear CMRs. 
Section 2 presents the pros and cons of the nonlinearity of CMRs, on which the present work is based. Section 3 applies the nonlinear color-to-metallicity conversion to the actual GC color distributions, and examines the inferred [Fe/H] distributions for M87 and M84 (\S\S\,3.1) and for the 100 early-type galaxies in the ACS Virgo Cluster Survey \cite[ACSVCS,][]{cote04} (\S\S\,3.2). Section 4 compares the inferred GC MDFs with the MDFs of spatially-resolved constituent stars of nearby galaxies (\S\S\,4.1 and \S\S\,4.2) and the MDFs from a simple chemical enrichment model of galaxies (\S\S\,4.3). Section 5 discusses the implications of our results on the color bimodality issues (\S\S\,5.1), addresses the question of whether the GC formation is coupled with the bulk formation of the stellar population of host galaxies (\S\S\,5.2), and finally presents our view on the formation and evolution of GC systems and their parent galaxies (\S\S\,5.3). 


\section{NONLINEARITY OF COLOR-METALLICITY RELATIONS: PROS~AND~CONS}

The core of this work is that the MDFs derived from GC optical color distributions 
are similar to those of constituent halo stars in galaxies. 
This work is based on the nonlinearity-CMR hypothesis (Paper I), 
that has been a target of dispute after its announcement. 
The issue is important enough that we devote a section to discuss it.

\subsection{Observed and Predicted Color-Metallicity Relations}

Simple linear conversion of photometric colors is
frequently used for estimating metallicities for large samples of extragalactic GCs.  
This is a reasonable first-order assumption for obtaining mean metallicities,
but for investigating the detailed structure of the MDF, including possible
subpopulations, the form of the CMR must be known to higher order.  
However, the best empirical color-metallicity calibrations currently available
\cite[e.g.,][]{peng06,mglee08,beasley08,sinnott10,woodley11,alves11}
exhibit notable observational scatter.
Moreover, compared to tens of thousands of GCs in a typical giant elliptical, the
calibration samples are still relatively small and sparsely populated at the
high-metallicity end.  
Larger samples of high-quality spectroscopic metallicities are needed to establish the
precise forms of CMRs.  
Such samples would implicitly include any correlations with age or other parameters, and would
provide strong constraints on the theoretical models.

In general, the slope of the dependence of a given photometric color on the
logarithmic metallicity [Fe/H] will change as a function of metallicity, 
and several recent studies have found departures from linearity. For instance, it has been known for decades that the color of the giant branch in Galactic GCs is a
nonlinear function of [Fe/H] \cite[e.g.,][]{michel84}.
\citet{richtler06}, using the observed $C-T_1$ versus [Fe/H] relation of \citet{harris02}, 
showed that the shape of the color distribution would differ from that of the MDF, 
and a non-peaked metallicity distribution could result in a bimodal color distribution.
\citet{mglee08} fitted nonlinear relations for $C-T_1$ color versus [Fe/H], and found
that the inferred MDF changed significantly depending on the adopted relation.
\citet{blakeslee10} fitted a quartic relation to the \citet{peng06} data and showed
that the empirical fit produces bimodal $g-z$ colors from unimodal MDFs.

As the observations have improved, so have the models, and these also
tend to predict nonlinear CMRs (e.g., Lee \etal\ 2002; Paper I; 
Cantiello \& Blakeslee 2007).  This is partly, but not totally, due
to improved modeling of the horizontal branch. Kissler-Patig \etal\ (1998a)
showed that the Worthey (1994) models predict a nonlinear relation between
[Fe/H] and $V-I$, with a wavy form qualitatively similar to that found empirically
by \citet{peng06}, although the inflection occurs at higher metallicity because
the colors of these models are generally too red at the high-metallicity end
(see Blakeslee \etal\ 2001).  This is interesting because the Worthey models do
not realistically model the horizontal-branch morphology, but treat it as a red clump near the
giant branch with a position that varies according to age.  The Lee
\etal\ (2002) model colors clearly showed nonlinear behavior as a function of
metallicity even without any horizontal-branch component, but the nonlinearity in the optical
colors was more pronounced with the horizontal branch.

Despite the advances, more work is needed, especially on the behavior of the horizontal branch in
extragalactic GC systems, as this is a complex multi-parameter problem.
Metallicity is the primary factor governing horizontal-branch temperature, and the transition
occurs in a nonlinear way at intermediate metallicities (e.g., Lee \etal\ 1994).
However, variations in other parameters, including age, helium content, and central
density, can create significant scatter in horizontal-branch morphology at a given metallicity
(e.g., Sandage \& Wildey 1967; Zinn 1980; Stetson \etal\ 1996, Sarajedini \etal\
1997; Buonanno \etal\ 1997; Sweigart \& Catelan 1998; see also the recent
discussions by Yoon \etal\ 2008; Gratton \etal\ 2010; Dotter \etal\ 2010).
There are also intricate intercorrelations among these parameters, as well as
correlations with GC mass.  For instance, the color-magnitude relation found among
the brightest GCs in external systems (e.g. Harris \etal\ 2006; Strader
\etal\ 2006; Mieske \etal\ 2006, 2010; Peng \etal\ 2009) is likely the result of a
mass-metallicity relation.  In Galactic GCs, the presence of an extreme blue
horizontal branch correlates strongly with GC mass (Lee \etal\ 2007), and is
likely related to the presence of helium-enhanced subpopulations (e.g., Norris
\etal\ 2004; Lee \etal\ 2005b; Piotto \etal\ 2005; Yoon \etal\ 2008; Han \etal\ 2009; Gratton
\etal\ 2010).  More massive GCs also have smaller half-light radii and higher
central densities (van den Bergh 1996).  Finally, there is some evidence for
correlations between age and metallicity in both the Galactic and extragalactic GC
systems (Puzia \etal\ 2005; Beasley \etal\ 2008; Mar{\'{\i}}n-Franch \etal\ 2009),
but the degree of correlation must depend on the formation history of the GC
system.

To summarize, the best current data indicate that optical CMRs 
for GC systems are nonlinear, and multiple sets of models support this
general result.  However, exactly how colors vary with metallicity 
depends on ages, the age-metallicity relations,  
and variations in other parameters such as $\alpha$-element and helium contents.  
Thus, a complete picture of the multi-band color-metallicity
behavior in extragalactic GC systems would include an understanding of the
interplay of the various stellar population parameters.  
Lacking more stringent
observational constraints, we consider only some simple, reasonable assumptions on
the standard stellar parameters, 
and for the present work, we
use our best predicted CMRs.

\subsection{Assessing the Evidence for Metallicity Bimodality}

\subsubsection{The Milky Way}
The GC system of the Milky Way Galaxy follows a bimodal MDF (Zinn 1985).  
The bimodality is confirmed by the radial number density profiles of metal-poor and metal-rich GCs 
and their orbital characteristics.
The metallicity distribution of GCs and field stars in our Galaxy are known
much more accurately than those in any other galaxy. 
For instance, the metal-poor GCs in the Milky Way have formed with an efficiency $\sim$ 20 times greater 
than the metal-rich GCs with respect to their associated stellar populations.
Although these were well known, there was little expectation that it would be a general
property of much larger GC systems in giant ellipticals.  Such galaxies contain
10-100 times as many GCs as the Milky Way, and reside predominantly in cluster
environments where they likely experienced very different evolutionary histories
(e.g., Peng \etal\ 2008); thus, the analogy with our own Galaxy was unclear.

In the context of the Toomre (1977) idea that elliptical galaxies are 
the remnants of dissipationally merged spirals,
Ashman \& Zepf (1992) discussed bimodal MDFs 
with a population of higher metallicity GCs forming in the major merger.
Nonetheless, given the simplicity of that model, and the advances in hierarchical structure
formation theory, it came as a surprise when most giant ellipticals exhibited
bimodal GC color distributions (interpreted as linearly reflecting metallicity),
and other scenarios were proposed to account for the bimodality (e.g., Forbes
\etal\ 1997; C\^ot\'e \etal\ 1998; Beasley \etal\ 2002; Kravtsov \& Gnedin \etal\ 2005;
Muratove \& Gnedin 2010).  However, the presumed bimodality of the MDF is taken as
an input constraint in these models, rather than being a clear
prediction (apart from perhaps the original Ashman \& Zepf scenario, which is
difficult to reconcile with the more complex assembly histories of ellipticals
found in cosmological simulations).  In light of this, it is worth reexamining the
direct evidence for bimodal MDFs in elliptical galaxies, and other galaxies with
comparably large GC systems.

\subsubsection{M31}

M31 is the largest galaxy in the Local Group and contains over 450 confirmed
GCs, three times as many as the Milky Way, and hundreds of additional candidates
(Galleti \etal\ 2004; Huxor \etal\ 2011; Caldwell \etal\ 2009, 2011).  Barmby \etal\ (2000) studied the
optical and near-IR color distributions for a large sample of M31 GCs.  Unlike in
most giant ellipticals, the colors did not appear bimodal, but Barmby \etal\ suggested
that errors in reddening and photometry could ``wash out'' the bimodality.  A KMM
analysis of the $V-K$ colors favored a double Gaussian model over a single Gaussian
with 92\% confidence ($<2\sigma$), but although the distribution appeared
asymmetric, it did not show two distinct components as in the Milky
Way.  The result was similar for the sample of $\sim160$ GCs with spectroscopic
metallicities compiled by these authors (see their Figure~19).

Galleti \etal\ (2009) presented a
homogeneous set of metallicities from Lick index measurements for 245 GCs.  Again,
they found that multiple Gaussian component models were favored because the MDF is
broad and asymmetric, but it lacks the two distinct metallicity peaks seen in the
Milky Way (see their Figure~15). 
when the sample was restricted to M31 GCs
with errors $<0.3$~dex, there was no appearance of bimodality.  Galleti
\etal\ (2009) conclude, ``The MD[F] of M31 GCs does not present any obvious
structure like the bimodality encountered in the GC systems of the MW [Milky Way].
Nevertheless, the distribution for M31 clusters does not seem to be well
represented by a single Gaussian distribution$....$ While clearly not
conclusive, the above analysis suggests that there may be actual structures in the
MD[F] of M31 GCs.''  

In the most recent study of the M31 GC MDF, \citet{caldwell11} present 
high signal-to-noise spectroscopic data on the M31 GC system.
Metallicities were estimated using a calibration of Lick indices with [Fe/H] provided by Galactic GCs. 
Although Caldwell et al. sample does not include many outer-halo metal-poor GCs 
that would increase the significance of the metal-poor side of the MDF,
the metallicity distribution of over 300 old GCs 
has a significant population of intermediate-metallicity GCs
and is not generally bimodal, 
in strong distinction with the bimodal Galactic GC distribution. 
The MDF shows a broad peak, centered at [Fe/H] = $-1$, 
possibly with minor peaks at [Fe/H] = $-1.4$, $-0.7$, and $-0.2$, 
suggesting that the GC systems of M31 and the Milky Way had different formation histories.
Given the complex accretion history of M31 (McConnachie
\etal\ 2009), it is not surprising that the M31 GC MDF would possess significant
structure, but the best current data do not present evidence for bimodality in the M31 GC system.
We note in passing that the Balmer absorption lines (H$\beta$, H$\gamma$ and H$\delta$)
theoretically have nearly the same response to horizontal-branch stars as optical broadband colors
in their index-metallicity relations (e.g., Lee \etal\ 2000; Chul \etal\ 2011, in prep.; S. Kim \etal\ 2011 (Paper IV)).
Remarkably, \citet{caldwell11}'s sample shows strong nonlinearity 
in the Balmer lines vs. metal line ($<$Fe$>$) relations (see their Figure 10),
and, as a result, exhibits clear Balmer strength bimodality (see their Figure 6)
which is a close analogy with optical color bimodality.

\subsubsection{Cen~A and the Sombrero}

The case for MDF bimodality is better in NGC\,5128 (Cen~A), an S0~pec
galaxy\footnote{``NED homogenized morphology,'' http://nedwww.ipac.caltech.edu/} at
the center of its own small group (Karachentsev \etal\ 2007).  Beasley \etal\
(2008) present spectroscopic metallicities for 207 GCs in this galaxy.  The
resulting MDF is skewed towards high metallicities and apparently has three closely
spaced peaks (see their Figure 5), in contrast to the two well-separated peaks in the
Milky Way.  This difference in MDF structure likely reflects the very different
accretion histories.  A very similar MDF was found by Woodley \etal\ (2010) in a
spectroscopic study of 72 NGC\,5128 GCs.  In this case, a unimodal distribution
provided statistically the best fit, but their [MgFe] index distribution was better
fitted with a double Gaussian model.  In both studies, the NGC\,5128 GC
metallicities and [MgFe] values lack the sharply bimodal appearance of the optical
colors, especially of $B-V$ and $V-I$ (Peng \etal\ 2004b).  However, Spitler \etal\
(2008) find that the optical--[3.6$\mu$m] IR color distributions for 146 NGC\,5128
GCs are distinctly bimodal, providing good evidence for MDF bimodality. 
They also found that similar data for a smaller sample of GCs in NGC\,4594 (the Sombrero) did not provide
significant evidence for or against MDF bimodality in that galaxy.
More recently, \citet{alves11} present a spectroscopic MDF for over 200 GCs in this galaxy, 
which is bimodal with peaks at [Fe/H] $\sim$ $-1.4$ and $-0.6$. 

\subsubsection{Giant elliptical galaxies}

Various studies of GC metallicities in giant ellipticals from spectroscopic and
near-IR/optical photometric data (Puzia \etal\ 2002, 2005; Cohen \etal\ 2003;
Strader \etal\ 2007; Hempel \etal\ 2007; Kundu \& Zepf 2007) are discussed in
detail by Blakeslee \etal\ (2010).  We summarize here and refer the reader to that
work for the full discussion.  

Despite the groundbreaking nature of many of these
studies, the spectroscopic results tend to be limited by the sample sizes ($<1\%$ of the
population), sample definitions, or sensitivity to the treatment of the data.
Thus, Blakeslee \etal\ (2010) conclude that the \textit{direct} evidence for
metallicity bimodality in giant ellipticals is weak.  
For example, the bimodality
reported by Strader \etal\ (2007) in the sample of 47 M49 GC metallicities derived
from the Cohen \etal\ (2003) Lick index measurements depended on the calibration
above solar metallicity.  Despite the very pronounced color bimodality, there was
no significant evidence for bimodality in the metallicities reported by Cohen
\etal\ (2003), which extended to higher metallicities, rather than forming a clump
near the solar value as in Strader \etal\ (2007).  Given the uncertainty in the
high-metallicity calibration (see discussion in Cohen \etal), and the sample
limitations (0.6\% of the GC population, observed with two spectroscopic masks), 
the issue remains unresolved for M49.
More recently, Foster \etal\ (2010) have studied the {Ca}\,\textsc{ii} triplet
(CaT) feature in a sample of 144 GCs in the Eridanus giant elliptical
NGC\,1407, which has prominent optical color bimodality.  In Galactic GCs, the CaT
index is linearly related to metallicity, at least for $\hbox{[Fe/H]}<-0.4$
(Armandroff \& Zinn 1988), and models indicate that it is very insensitive to age
(Vazdekis \etal\ 2003).  Foster \etal\ (2010) find that bright GCs near the peaks
of the color distribution have very similar CaT strengths, indicating very similar
metallicities despite the wide separation in color space. The distribution of
CaT-derived metallicities does not appear bimodal, but because of its asymmetry, it
is better fitted by a double Gaussian model at the $2\sigma$ level.  However, if
this result were interpreted as MDF bimodality, then the components would differ
significantly in amplitude, width, and position from those implied by the optical
colors; i.e., it would be a different bimodality.  Foster \etal\ remark that if the
metallicities are taken at face value, then the very different color and
metallicity distributions could be reconciled by a nonlinear CMR 
causing a unimodal MDF to appear bimodal in color space.

In a recent near-IR/optical photometric study, 
\citet{kundu07} presented the $I-H$ color distribution of 80 GCs in M87,
which shows bimodality. 
More recently,
\citet{chies10,chies11a,chies11b} present (optical -- near-IR) colors for the GC systems in 14 early-type galaxies,
and find that the bimodality becomes less evident in $g-K_s$ if compared to $g-z$ 
and even less pronounced in $z-K_s$. \citet{chies10} point out that
the disappearance of bimodality in these colors while evident in the optical $g-z$ color 
could be attributed to a nonlinear-CMR effect 
although the observational uncertainties could also account for it. 
Finally, we note that our new studies combining ACS and WFC3/IR data 
for NGC\,1399 in the Fornax galaxy cluster (Blakeslee \etal\ 2011, in prep.) 
and Subaru/MOIRCS near-IR and CTIO optical data 
for M60 and NGC 4365 in the Virgo (S. Kim \etal\ 2011, in prep.)  
find independently that the $I-H$ and $I-K_s$ color distributions are not significantly bimodal, despite
the strong bimodality in the optical colors of the same sample.

Overall, the available data on the MDFs of GCs in giant ellipticals are at best
ambiguous.  In cases where spectroscopy -- the more {\it direct} measures of metallicity than colors -- suggest bimodality,
it is much less apparent than the dramatic double-peaked histograms of colors for the same galaxies (e.g., Peng \etal\ 2006).  
Thus, at least \textit{some} of the observed
color bimodality is likely due to nonlinear behavior of colors with metallicity.  As
discussed above, there is empirical evidence for such
nonlinearity having a form that tends to produce bimodal color
distributions.  Pipino \etal\ (2007), considering the results of Puzia \etal\ 
(2005), also concluded that color-metallicity nonlinearity would help significantly
in reconciling the photometric and spectroscopic data.
The question of the relative importance of color-metallicity nonlinearity and
metallicity bimodality in producing the observed GC color distributions remains open.
In the meantime, besides the necessity of better spectroscopic samples, it is worthwhile 
to explore the possibility of deriving MDFs from the observed colors
under the assumption of nonlinear CMRs.
The following sections use the latest stellar population models to invert the colors for very large
photometric samples and to examine the implications of the resulting MDFs.


\section{COLOR AND METALLICITY DISTRIBUTIONS OF EXTRAGALACTIC GLOBULAR CLUSTER SYSTEMS}

Our main objective is to investigate the MDFs for GC systems 
when inferred from the nonlinear CMRs. 
Compared to the metallicity-to-color conversion shown in Paper I,  
the inverse-conversion from colors to metallicity is more susceptible to  
the inevitable incompleteness of current population synthesis models.
With a theoretical CMR that is somewhat incorrect in the color direction, for example, 
the metallicity-to-color conversion will still give color distributions with the correct shape,
but the inverse-conversion will yield erroneous metallicity distributions. 
Moreover, the inverse-conversion from colors to metallicity may be hampered by 
the varying observational uncertainties depending on the colors of interest.
With these caveats in mind, however, careful inverse-conversions may shed light on 
the structure of the GC MDFs, including possible subpopulations. 
In this section, we apply the transformations of Paper I to the color distributions of GC systems 
and present their inferred MDFs
for M87 and M84 (\S\S\,3.1) and for 100 early-type galaxies imaged in the ACSVCS (\S\S\,3.2).

\subsection{Globular Cluster Systems in M87 and M84}

In this section, we present the results of our multiband photometry for GC systems of M87 (NGC 4486) and M84 (NGC 4374). 
We have selected M87 and M84, giant elliptical galaxies in the Virgo cluster, because they both have GC systems with confirmed color bimodality in $g-z$ and are among very few elliptical galaxies with deep $u$-band observations available.  
We refer the reader to Yoon \etal\ (2011, hereafter Paper II) for greater details on the multiband photometry of the M87 GC system in the context of nonlinear CMRs. The archival F336W images of {\it HST}/WFPC2 and {\it HST}/WFC3 were used to obtain $u_{F336W}$ for GC candidates in M87 and M84, respectively. Our $u_{F336W}$-band catalogs were matched with ACS/WFC $g_{F475W}$- and $z_{F850LP}$-band photometry of Jord\'{a}n et al (2009). 
We hereafter refer to  $u_{F336W}$, $g_{F475W}$, and $z_{F850LP}$ mags as $u$, $g$, and $z$, respectively.
Jord\'{a}n et al (2009) selected {\it bona-fide} GCs with their magnitudes, $g-z$ colors, and sizes. We further employed color cuts in the $u$-band colors to filter out contaminating sources, especially background star-forming galaxies. We used 591 GCs in M87 and 306 GCs in M84 that have reliable $u$, $g$, and $z$ measurements in common. The samples are $u$-band limited.

The merit of the multiband observations is clear: Since the form of CMRs hinges on which color is used, the shape of the color distributions varies significantly depending on the colors in use. Hence a comparative analysis of the GC MDFs that are independently obtained from distributions of different colors will put the nonlinearity hypothesis to the test, as proposed in Paper I and Paper II. Among other optical colors, the $u$-band related colors (e.g., $u-g$ and $u-z$) are theoretically predicted to exhibit the most distinctive CMRs from other preferred CMRs (e.g., for $g-z$), and thus the most adequate to the task. Furthermore, the $u$-band colors are significantly less affected by the variation in the horizontal-branch mean temperature, having less inflected, ``smooth'' CMRs than $g-z$ for given ages. Therefore, for instance, the conversion from $u-g$ color distributions to MDFs via the ($u-g$)-[Fe/H] relation should be more straightforward than the case of $g-z$.   
The reason why the CMRs for $u$-band colors are less inflected than the $g-z$ CMR is two-folded: 
($a$) the integrated $u$-band colors of main-sequence and red-giant-branch stars are smoother functions of metallicity 
compared to $g-z$, and ($b$) the $u$-band colors are less sensitive to the horizontal-branch temperature variation,
which is due to the fact that the blueing effect of the optical spectra with increasing horizontal-branch temperature 
is held back by the Balmer discontinuity where the $u$-band is located (Yi \etal\ 2004).
Such properties make the $u$-band colors good metallicity indicators for a wide range of age, 
and the $u$-band color distributions are expected to be
significantly different from distributions of other optical colors such as $g-z$, $V-I$, and $C-T_1$.
See Paper II for detailed discussion on the $u$-band colors as a tool to probe the nonlinearity of CMRs.

Figure 1 shows the observed color distributions of the M87 GC system from $u$, $g$, and $z$ photometry and the inferred MDFs. 
First, Figures 1$a$ and 1$d$ present the $g-z$ vs. $u$ and $u-g$ vs. $u$ diagrams, respectively. 
Figures 1$b$ and 1$e$ present the $g-z$ and $u-g$ color distributions, respectively. 
The $g-z$ distribution of the M87 GCs unambiguously displays two peaks around $g-z$ = 1.0 and 1.4. 
In contrast, the $u-g$ distribution for the identical sample does not appear to have clear bimodality. 
One may argue that the larger observational uncertainties in $u$-band weaken the bimodality. 
Table 3 shows that the typical photometric error of $u-g$ is 
2.2 times larger than that of $g-z$ for the entire sample,
but at the same time the ranges spanned by the colors 
are $\Delta$($g-z$) = 1.1 mag and $\Delta$($u-g$) = 2.1 mag,
that is, the baseline of $u-g$ is 1.9 times longer than that of $g-z$.
As a result, the relative sizes of error bars are ($g-z$ : $u-g$) = (1.0 : 1.2). 
In a relative sense, the errors in the two colors are quite comparable to each other.
Moreover, Paper II shows that the $u-z$ color, which has {\it smallest} relative errors, still exhibits weaker bimodality in the color distribution compared to the $g-z$ distribution.
It is, therefore, not likely that bimodality in the $u-g$ distribution of M87 GCs
is simply blurred by larger observational errors in the $u$-band.

The variation in the histogram shape for different colors may suggest that the form of the CMRs varies depending significantly on the colors in use. This is shown in Figures 1$c$ and 1$f$, along with our predictions from the Yonsei Evolutionary Population Synthesis (YEPS) model\footnote{The models in this study are constructed using the Yonsei Evolutionary Population Synthesis (YEPS) code. The YEPS model generates ($a$) synthetic color-magnitude diagrams for individual stars \cite[see, e.g.,][]{lee94,lee99,lee05b,rey01,yoon02,yoon08,han09} and ($b$) synthetic integrated spectra for colors and absorption indices of simple and composite stellar populations \cite[see, e.g.,][]{lee05a,par97,lee00,rey05,rey07,rey09,kaviraj05,kaviraj07a,kaviraj07b,kaviraj07c,ree07,yoon06,yyl09,yoon09,spitler08,mieske08,choi09,cho11,yoon11}.
One of the main assets of our model is the consideration of the systematic variation in the mean color of horizontal-branch stars as functions of metallicity, age, and abundance mixture of stellar populations.
The standard YEPS model employs the Yonsei-Yale stellar evolution models (Y. Kim \etal\ 2002; Han \etal\ 2011, in prep.) and the BaSeL flux library (Westera \etal\ 2002). The spectro-photometric model data of the entire parameter space are available at http://web.yonsei.ac.kr/cosmic/data/YEPS.htm.} (Chung et al. 2011; Yoon et al. 2011, {\it in prep.}).  In Figure 1$c$, the $g-z$ colors are shown as a function of [Fe/H] for GCs in the Milky Way (open circles), and M49 and M87 (filled circles and triangles). 
The references to the observed data used in the relations are summarized in Table 1. The fifth-order polynomial fit to our model data for 13.9-Gyr GCs is overlaid (thick solid line, see Table 2). For a comparison, the straight grey line represents the linear least-squares fit to the data. 
Figure 1$f$ is the same as Figure 1$c$, but for $u-g$ color. Open circles, blue and red filled squares represent GCs in the Milky Way, M87, and NGC 5128, respectively. 
The $u-g$ colors of the GCs in the Milky Way and NGC 5128 were obtained from their $U-B$ colors via the equation, ($u-g$) = 1.014 ($U-B$) + 1.372, derived from model data for synthetic GCs with combinations of age (10 $\sim$ 15 Gyr of 1 Gyr intervals) and [Fe/H] ($-2.5$ $\sim$ 0.5 dex of 0.1 dex intervals). 
$U$- and $B$-passband are relatively close to $u$- and $g$-passband, respectively, and thus $U-B$ has responses to the horizontal-branch morphology in a way that is very similar to $u-g$. 
Therefore, $U-B$ are a good proxy to $u-g$, 
and the $U-B$ vs. $u-g$ relationship is best described by a linear fit over a range of ages and metallicities (Table 1). 
Guided by the model, the form of the observed CMRs appears to vary from the ($g-z$)-[Fe/H] relation to ($u-g$)-[Fe/H].  One may argue, however, that current data appear to be fit both by the theoretical nonlinear relations and the empirical straight relations, as there is only a weak indication that the modeled relations are actually better fits to the $g-z$ vs. [Fe/H] and $u-g$ vs. [Fe/H] data. Given the current level of observational accuracy and inhomogeneity of the data, the purpose of our study is not to determine the exact shape of color-metallicity relationships, but to investigate the consequences and implications of the possible nonlinear transformations.

We now consider the inferred GC MDFs. Figures 1$g$ and 1$h$ show the MDFs for the M87 GC system. On the one hand, Figure 1$g$ presents the GC MDFs converted from $g-z$ (red histogram) and $u-g$ (blue histogram) colors that are based on the traditional {\it linear} color-to-metallicity conversion (thin grey lines in Figures 1$c$ and 1$f$), and thus are just replicas of their color histograms shown in Figures 1$b$ and 1$e$. Note that both the overall shape and the peak positions do not appear to agree between the GC MDFs from the two colors. On the other hand, Figure 1$h$ presents the GC MDFs converted from $g-z$ and $u-g$ colors that are based on the improved {\it inflected} relationship between color and metallicity (thick black lines in Figures 1$b$ and 1$e$). In contrast to Figure 1$g$, the inferred GC MDFs in Figure 1$h$ are modified drastically to have a strong metal-rich peak with a metal-poor tail. 
The two histograms in Figure 1$h$ are more consistent with each other in terms of their overall shape and peak positions than those shown in Figure 1$g$. 
We note that our stellar population models show that, for given input parameters, the {\it absolute} quantities of output are rather subject to the choice of model ingredients such as stellar evolutionary tracts and model flux libraries; the different choices can result in up to $\sim$ 0.2 mag $g-z$ and $u-g$ variation among models and the inferred [Fe/H] values accordingly (up to $\sim$ 0.5 dex). 
Hence, one should put more weight on the {\it relative} values of inferred GC MDFs, 
i.e., the overall morphology of the MDFs and their unimodality.  
We, however, wish to emphasize that the typical GC MDF shape is obtained invariably from different colors, i.e., $g-z$ and $u-g$ for M87 GCs.

Figure 2 is the same as Figure 1, but for the GC system in M84. First, Figures 2$a$ and 2$d$ present the $g-z$ vs. $u$ and $u-g$ vs. $u$ diagrams for the M84 GCs. Figures 2$b$ and 2$e$ present the $g-z$ and $u-g$ color distributions, respectively. 
As for the possible role of observational uncertainties in weakening bimodality of the $u-g$ color distribution,
the typical photometric error of $u-g$ is 1.7 times larger than that of $g-z$ for the entire sample (Table 3), 
but at the same time the ranges spanned by the colors 
are $\Delta$($g-z$) = 1.1 mag and $\Delta$($u-g$) = 1.8 mag,
that is, the baseline of $u-g$ is 1.6 times longer than that of $g-z$.
As a result, the relative sizes of error bars are ($g-z$ : $u-g$) = (1.0 : 1.1). 
In a relative sense, the errors in the two colors are comparable.
It is, therefore, not likely that bimodality in the $u-g$ distribution of M84 GCs
is simply blurred by larger observational errors in the $u$-band. 
Figures 2$c$ and 2$f$ are the same as Figures 1$c$ and 1$f$, respectively. Finally, Figures 2$g$ and 2$h$ show the inferred MDFs for the GC systems in M84. 
Again, the inferred GC MDFs in Figure 2$h$ have a strong metal-rich peak with a metal-poor tail. 
The two histograms in Figure 2$h$ show a better agreement with each other in terms of their overall shape and peak positions than those in Figure 2$g$.

In this section, we have obtained multi-band colors of GCs in the two representative giant elliptical galaxies, M87 and M84, and examined their color and metallicity distributions. We have found that the distributions of different colors can be transformed into unimodal metallicity distributions that are strongly peaked with a broad metal-poor tail.
The implications of the typical shape of the inferred GC MDFs and its similarity to those from chemical evolution models and field-star observations (see \S\,4) will be discussed in \S\,5. 
We note, however, that the similarity itself between the GC MDFs from multiband colors does not necessarily represent evidence that the model is correct. Whether the similar MDFs from various colors can be taken as evidence for the nonlinear-CMR scenario for the color bimodality is a sufficiently involved issue and fully explored in Paper II.

\subsection{Globular Cluster Systems in the ACS Virgo Cluster Survey}

Motivated by the findings above in \S\S 3.1 for the individual galaxies and to avoid possible small-number statistics, we now benefit from the 100 early-type galaxies in the ACSVCS \citep{cote04,peng06,jordan09}. We apply the color-to-metallicity transformation scheme to $\sim$10,000 GCs in ACSVCS, the largest and most homogeneous photometric database of extragalactic GCs currently available. Figure 3  presents the observed color distributions and inferred MDFs of GC systems in bins of host galaxy luminosity. In Figure 3$a$, we show the observed color histograms of GC systems for seven bins of host galaxy magnitude. The data are the same as in Figure 6 of \citet{peng06} and here we listed the data in Table 4. The histograms are normalized by the GC number at their blue peaks, and multiplied by constants, $C$, for clarity. The magnitude bins are 1 mag wide and extend from $M_B$ $\simeq$ $-$21.5 ($-$22 $\leq$ $M_B$ $<$ $-$21, red, $C$ = 1.0) to $\simeq$ $-$15.5 ($-$16 $\leq$ $M_B$ $<$ $-$15, purple, $C$ = 0.4). A Gaussian kernel of $\sigma$($g-z$) = 0.05 is applied. Clearly, the histograms appear bimodal or asymmetric across the entire luminosity range, with all GC systems containing blue peaks and with more prominent red peaks in brighter hosts. 

A close scrutiny of Figure 3$a$ reveals the tendency of the dip positions (and blue peak positions) in the color histograms to become progressively bluer as the host luminosity decreases.  In the context of nonlinear CMRs, this can be explained if GCs in fainter galaxies are slightly younger than those in brighter galaxies. This is because at younger ages, the blue horizontal branch develops at lower metallicity \citep{lee94,yoon08,dotter10} and, as a consequence, the predicted colors of the quasi-inflection points along the CMR move systematically towards the blue at younger ages. This effect is demonstrated in Figure 3$b$ by an example set of the YEPS model predictions with $\Delta$$t$(brightest--faintest) = 3 Gyr. In this example, ages range from 10.5 Gyr (purple) to 13.5 Gyr (red) by equal intervals of 0.5 Gyr. The YEPS $g-z$ data for 9 Gyr to 14 Gyr by steps of 0.5 Gyr are given in Table 2. The thin straight dotted line represents the linear least-squares fit to the data shown in Figures 1$c$ and 2$c$.

Figures 3$c$\,--\,3$f$ present the results of the four different color-to-metallicity transformations. Like the color distributions in Figure 3$a$, the inferred GC MDFs are displayed for seven bins of $M_B$ from $M_B$ = $-$22 to $-$15 in steps of 1 mag. To obtain these MDFs, we converted the color of each GC into [Fe/H] using the ($g-z$)-[Fe/H] relations from the YEPS model. Each panel makes a different assumption on the systematic age sequence from the faintest host bin to the brightest. The modeled ages for host luminosity bins are shown in the insets of Figures 3$c$\,--\,3$f$, with age differences, $\Delta$$t$ = 0, 1, 2, and 3 Gyr, respectively, between the faintest ($M_B$ $\simeq$ $-$15.5, purple) and brightest ($-$21.5, red) bins. The brightest, oldest ($M_B$ $\simeq$ $-$21.5, red) bin is set to be 13.5 Gyr. The color distribution shown in Figure 3$a$ was transformed to MDFs via the model CMRs of the corresponding ages. 

Figure 3$c$ presents the case in which the age is assume to be constant at 13.5 Gyr regardless of host luminosities between $M_B$ $\simeq$ $-$15.5 (purple) and $-$21.5 (red). The GC MDFs for the luminous ($M_B$ $<$ $-$17) host bins (the first to fifth brightest bins) in particular are strongly peaked with a broad metal-poor tail. As Paper I suggested, the strong bimodality seen in the GC color distribution of luminous galaxies is not evident in the MDF once transformed by their wavy CMR. The MDFs in the faintest two bins, however, have broad peaks. This is likely because the less luminous galaxies primarily have blue GCs and their colors are largely bluer than the inflection point in the CMR under the constant age assumption, i.e., $\Delta$$t$(brightest -- faintest) = 0 Gyr. Paper I mentioned that the position of the inflection is bluer for younger stellar populations and so if fainter galaxies host GCs younger than brighter counterparts, then a different CMR may apply. However, in this naive use of the single CMR, not all inferred [Fe/H] distributions appear to be unimodal.

Comparative analysis shows that the non-zero age difference of $\Delta$$t$(brightest -- faintest) = 1\,$\sim$\,3 Gyr (Figures 3$d$\,--\,3$f$) results in the GC MDFs for {\it all} the host luminosity bins that fit better with skewed Gaussian distributions with metal-poor tails. The grey dotted histogram in each panel represents the inferred MDF for the brightest (i.e., $M_B$ $\simeq$ $-$21.5) bin based on the simple straight fit in Figure 3$b$, and is intended for comparison to the red solid MDF. Obviously, the strong bimodality seen in the color distributions of luminous galaxies is no longer evident in the MDFs, once transformed via the inflected color-metallicity relationship. The inferred GC MDFs for all seven host luminosity bins have the same characteristic shape -- being sharply peaked with a broad metal-poor tail -- across three orders of magnitude in the host galaxy mass. In addition, the mean [Fe/H] and peak position of the GC MDFs are a strong function of the host luminosity, in the sense that, for brighter host galaxies, the mean [Fe/H] increases and the peak gets redder.

Compared to their relative ages, the absolute ages of GC systems are still less certain. To test the robustness of the result shown in Figure 3 against different absolute ages, Figure 4 makes differing assumptions on the age sequence from the faintest host galaxies to the brightest. In this case, the center bin, i.e., the fourth brightest ($M_B$ $\simeq$ $-$18.5, green) bin, is set to be 13 Gyr. The modeled ages for the host luminosity bins are shown in the insets of Figures 4$c$\,--\,4$f$, with age differences of $\Delta$$t$ = 0, 1, 2, and 3 Gyr, respectively, between the faintest ($M_B$ $\simeq$ $-$15.5, purple) and brightest ($-$21.5, red) bins. In Figure 4$b$, another example set of our model prediction with $\Delta$$t$ = 3 Gyr is shown. In this example, ages range from 11.5 Gyr (purple) to 14.5 Gyr (red) in equal intervals of 0.5 Gyr (Table 2). Figures 4$c$\,--\,4$f$ show that the strong bimodality is no longer evident in the inferred MDFs, and the typical shape of the GC MDFs does not depend upon differing age assignment from the faintest host galaxies to the brightest. It also holds true that the peak positions of the GC MDFs are a strong function of the host luminosity in that, for brighter host galaxies, the mean [Fe/H] increases and the peak gets redder. 
We also tested the stability of the results against the putative age dispersion in the examined data. The typical form of the GC MDFs persists with up to $\sigma_t$ $\simeq$ 2.5 Gyr, where the age spread among GCs in a single host luminosity bin is parameterized by a Gaussian dispersion, $\sigma_t$. The data for the inferred GC MDFs in Figures 3$d$ and 4$d$ (they are identical) are given in Table 5.

If one could assume that the intrinsic shapes of the GC MDFs are similar in all host luminosity bins, our results might give a tantalizing hint of varying ages among GC systems in that GCs in more luminous parents are older than GCs in fainter galaxies. However, there is yet no observational support that the assumption is valid. 
Moreover, the {\it true} shape of CMRs suggested in this study is still unproven. 
Hence, the purpose of our simulations should be to determine neither the absolute age of the GC systems nor their exact pattern of age sequence from the faintest bin to brightest. Nevertheless, the possible age sequence appears not inconsistent with the ``galaxy downsizing'' picture \cite[e.g.,][]{cowie96} in which brighter galaxies are observed to form earlier. If confirmed, this can be regarded as another indication of the common characteristics shared by stellar populations of GCs and halo field stars (see \S\,4). 

\section{COMPARISON OF GC MDF'S TO THOSE OF HALO FIELD STARS AND GALAXY CHEMICAL EVOLUTION MODELS}

We find that the bimodal GC color distributions commonly observed in luminous early-type galaxies are transformed into unimodal metallicity distributions that are strongly peaked with a broad metal-poor tail. Our sample includes GC systems in M87 and M84 (\S\S\,3.1) and in the ACSVCS galaxies (\S\S\,3.2). A key to comprehending the connection between GCs and halo field stars in galaxies is a direct comparison of their MDFs. This section compares the inferred GC MDFs to those of resolved field stars in nearby early-type galaxies (\S\S\,4.1 and \S\S\,4.2). Also, the GC MDFs are compared to MDFs produced by chemical evolution models of galaxies (\S\S\,4.3).  

\subsection{Comparison of ACSVCS GCs to Spatially-resolved Halo Field Stars in Nearby Elliptical Galaxies}

The necessity of obtaining photometry of spatially-resolved field stars limits us to nearby galaxies. There are several nearby, relatively massive elliptical galaxies whose stellar MDFs have been measured, and our inferred GC MDFs can directly be compared to such stellar MDFs. Figure 5 gives a comparison of the MDFs of the ACSVCS GCs to those of resolved stars of nearby elliptical galaxies \citep{harris02,rejkuba05,harris07a,harris07b,bird10}. The stellar MDFs of individual galaxies were obtained from color-magnitude diagrams of red-giant stars whose colors are highly sensitive to their metallicities. 

In Figure 5, we plot the GC MDF shown in Figures 3$d$ and 4$d$ (they are identical). The data for the plot are presented in Table 5. The figure shows that the GC MDFs are similar to the MDFs of resolved constituent stars of nearby elliptical galaxies. In particular, the typical shape of the GC MDFs (colored solid line in each panel), characterized by a sharp peak with a metal-poor tail, is remarkably consistent with those of field stars in nearby galaxies. By contrast, GC MDFs obtained using the simple linear [Fe/H] vs. $g-z$ relation (grey dotted lines) do not agree with the stellar MDFs. We note that the metallicity spread of GCs tends to be broader than that of stars. As indicated in Figures 3$c$\,--\,3$f$ and Figures 4$c$\,--\,4$f$, the GC MDFs, on the whole, are broader than the stellar MDFs. This is likely due to the fact that each color histogram of the ACSVCS consists of GCs belonging to 3 $\sim$ 20 diverse host galaxies, giving an ensemble character of each host luminosity bin. Moreover, the inferred MDFs become broader as the observational uncertainty in color is propagated to metallicity space. 

In addition to the noticeable similarities found in the shape of MDFs between GCs and stars, they share a common feature in that their mean metallicity increases with increasing host luminosity. As a result, the peak positions of GCs and stars are roughly coincident in each luminosity bin --- at [Fe/H] $\simeq$ $-$1.0 (faint hosts) and $-$0.5 (bright hosts). This, however, should be taken with great caution because the mean colors of both GCs \citep{dirsch03,jordan04,tamura06,mglee08} and field stars \citep{harris02,rejkuba05} depend significantly on the sampled radial location in a galaxy. Their mean metallicities gradually decrease with projected radius, as seen in Figures 5$d$ and 5$e$ for the outer- and inner-halo stars of NGC 5128, respectively. Moreover, observations show that the mean GC color is bluer than stars within an elliptical galaxy as a whole \citep{peng06}. More importantly, GCs are on average bluer than the unresolved light of the galaxies at the same radii \cite[e.g.,][]{forte81,strom81,jordan04,tamura06,mglee08,forte09}. For old stellar populations such as those in elliptical galaxies and GCs, the observed bluer colors imply lower metallicities in all stellar models, thus leading to the conclusion that GCs are typically more metal-poor than field stars in a galaxy. Hence it may just be a coincidence that the GC MDFs line up with the field star MDF of the nearest elliptical galaxy. Despite these caveats, the peak and width of the MDFs of GCs and field stars are similar enough to suggest that both were formed in the same events, which built the major part of the galaxies. 

It should be addressed that the [Fe/H] histogram for NGC 3379 stars (Figure 5$c$), despite the overall resemblance to the compared GC MDF, shows an excess of metal-poor stars with a bump at [Fe/H] $\simeq$ $-$1.2. \citet{harris07b} found the metal-poor stellar halo in NGC 3379 and provided an explanation for why their earlier studies only detected a metal-rich component in the same galaxy. The NGC 3379 {\it HST} field is at a distance of 12 effective radii, which is more than twice as far as the equally small fields of view for NGC 3377 (Figure 5$b$) and NGC 5128 (Figures 5$d$ and 5$e$). Studies of the stellar MDFs of large galaxies have been performed photometrically and concentrate on more central regions of the galaxy where the metal-rich field star population is concentrated. The radial bias seemed to under-sample metal-poor stars in the earlier studies. With the recent evidence of metal-poor halo stars emerging at great galactocentric distances, one should not assume that the stellar MDFs are unimodal at all positions. Therefore, this suggests that our interpretation should be more applicable to their main, inner parts of galaxies, less to the remote outskirts. 
 
\subsection{Comparison between Spatially-resolved Halo Field Stars in Nearby Elliptical Galaxies and their Own GC systems}

As envisaged by the NGC 3379 case above, each galaxy has its own evolutionary history. So a more direct way to assess the similarity between GCs and stars is to compare between the MDFs of spatially-resolved field stars in a galaxy and its {\it own} GC system. There are four galaxies (M87, NGC 5128, NGC 3377, and NGC 3379) in Figure 5, for which both the GC MDF and the stellar MDF are currently available. Figure 6 gives the comparison for M87 (the top row), NGC 5128 (the second row), NGC 3377 (the third row), and NGC 3379 (the bottom row). The stellar MDFs (grey histograms) are identical to those in Figures 5$b$, 5$c$, 5$d$, and 5$f$. The left-hand panels present the inferred GC MDFs (empty histograms) based on the traditional linear color-to-metallicity transformations. By contrast, the right-hand panels show the inferred GC MDFs (empty histograms) based on the inflected relations from the YEPS model (Table 2). 

Firstly, for M87, we exploit $g-z$ colors from the ACSVCS \citep{peng06,jordan09} to derive the GC MDF. 
Since the inferred GC MDFs shown in Figure 1$h$ are based on the $u$-band limited sub-sample, 
the ACSVCS $g$- and $z$-band data are more representative GC sample.
Figure 6$a$ shows that the MDF (empty histogram) for 1745 GCs obtained using the simple linear [Fe/H] vs. $g-z$ relation (shown in Figure 1$c$) exhibits a fundamental difference from the stellar MDF \citep{bird10} measured in a similar region of the same galaxy. By contrast, in Figure 6$e$, the strong bimodality is no longer present once transformed by the inflected CMR (Table 2), and the GC MDF is very similar to that of the brightest host bin in the ACSVCS. As a consequence, the MDFs of GCs and field halo stars in M87 are similar in shape and line up remarkably well with each other.

Secondly, for NGC 5128, \citet{peng04a,peng04b} presented the CTIO Blanco 4-m $U$-, $B$-, $V$-, and $I$-band photometry of the GC system. We use the $B-I$ color distribution to derive the GC MDF because $B-I$ is a reasonable substitute for $g-z$. However, our result on NGC 5128 is not affected by the choice of colors. Figure 6$b$ shows the MDF (empty histogram) for 210 GCs that has strong bimodality. On the other hand, in Figure 6$f$ we converted the $B-I$ colors into [Fe/H] using the YEPS relation (Table 2). The strong bimodality seen in Figure 6$b$ is not evident in the MDF, and the strongly peaked GC MDF with a broad metal-poor tail is in reasonably good agreement in shape with the stellar MDF \citep{rejkuba05}. 

An unavoidable uncertainty in this comparison is that the field stars are sampled from only one or two projected location(s), whereas the GCs cover the wider halo. For M87, the stars were sampled in an inner region (R $\simeq$ 10 Kpc) and the GCs were in the inner halo ($\lesssim$ 10 Kpc). For NGC 5128, the stars were sampled in an inner region (R $\simeq$  20 and 30 Kpc) and the GCs were in the entire halo of the galaxy. This may partly explain the similarity of the peak positions between stellar and GC MDFs for M87, and the dissimilarity for NGC 5128. Another warning for NGC 5128 is that, although the majority of its GCs ($\sim$ 90 \%) are known to be old ($>$ 10 Gyr) \citep{beasley08}, there should be a certain portion of young GCs for which one needs to apply a different color vs. metallicity relation to derive the GC MDF. Nonetheless, despite these sources of uncertainty, it is engrossing that the inferred GC MDFs based on the inflected color-to-metallicity transformations gives better matches with the field star MDFs of M87 and NGC 5128, compared to those based on the traditional linear relations. 

Thirdly, for NGC 3377, \citet{cho11} presented the {\it HST}/ACS $g$- and $z$-band photometry of the GC system as part of a deep imaging study of 10 early-type galaxies in low-density environments. Figure 6$c$ shows that the MDF for 157 GCs derived from the simple linear [Fe/H] vs. $g-z$ relation has strong bimodality. On the other hand, in Figure 6$g$ we converted the $g-z$ color into [Fe/H] using the YEPS relation. The strong bimodality seen in Figure 6$c$ is not evident in the MDF, and the sharply peaked MDF with a broad metal-poor tail is in reasonably good agreement in shape with the stellar MDF \citep{harris07a,harris07b}. 

Lastly, for NGC 3379, \citet{whitlock03} and \citet{rhode04} carried out wide-field photometry of GCs. They verified earlier results that its GC population is quite small. \citet{harris07b} derived the GC MDF from Rhode \& Zepf (2004)'s $B-R$ histogram for 36 GCs with an improved empirical relation, [Fe/H] = 3.13 ($B-R$)$_0$ -- 5.04. Figure 6$d$ shows the GC MDF, which is highly concentrated toward the metal-poor side. If divided at [Fe/H] = $-$1.2 where the metal-poor bump of the filed-star MDF is located, the metal-poor GCs outnumber the metal-rich GCs by 25 to 11. As a result, for the metal-rich half of the MDF, the numbers of GCs are much too small. By contrast, Figure 6$h$ displays the GC MDFs based on the YEPS relation (Table 2). With the GC MDFs of only 36 clusters, it is not yet clear whether the MDFs of GCs and stars have the same shape. It is interesting to note, however, that the GC metallicities are now more evenly distributed, and the proportions of the two metallicity subgroups, when divided at [Fe/H] = $-$1.2, became more comparable (21\,:\,15) than those in Figure 6$d$ (25\,:\,11). Interestingly, the inferred GC MDF in Figure 6$h$ seems very similar to the NGC 3379 field-star MDF in the western half (farther from the galaxy center) of the ACS/WFC field (shaded histogram in Figure 14 of Harris \etal\ 2007b) 

\subsection{Comparison to Chemical Enrichment Models of Galaxies}

The new development in linking the observed GC colors to their intrinsic metallicities leads to GC MDFs that are strikingly similar in shape to MDFs of resolved field stars in nearby elliptical galaxies. Both the inferred GC MDFs and halo stellar MDFs are characterized by a sharp peak with a metal-poor tail. In this section, we proceed to compare the inferred GC MDFs to chemical enrichment models of galaxies.

In Figure 7 we compare the inferred GC MDFs for the second brightest ($M_B$ $\simeq$ $-$20.5, orange) and the brightest ($M_B$ $\simeq$ $-$21.5, red) galaxy luminosity bins with the simple closed-box model of chemical evolution \cite[e.g.,][]{pagel75}. We plotted the MDFs for this simple model using yields of [Fe/H] = $-$0.685 and $-$0.505, respectively. With the improved nonlinear relationship between color and metallicity, the general shape of GC MDFs is in remarkable agreement with that of galaxy chemical enrichment models. In contrast, the dotted line in each panel represents the inferred GC MDF for the corresponding bin based on the simple straight fit to the data.  

Despite their resemblance in the general shape, the inferred GC MDFs have fewer metal-poor GCs than the simple chemical model. This is akin to the ``G-dwarf problem'' in the solar neighborhood \cite[e.g.,][]{vandenbergh62,schmidt63}. We note that the width of the chemical evolution model can be changed using various kinds of gas infall and stellar feedback, and an accreting-box model of chemical evolution yields narrower distributions that provide a better match with the inferred GC MDFs. For instance, \citet{harris02} remarked that the halo of NGC 5128 ($M_B$ = $-$20.9) also suffers from the G-dwarf problem in that it lacks metal-poor stars compared to the simple model. They were able to fit the field star MDF using an accreting-box model of chemical evolution, producing a narrower distribution that is also a better match to the inferred GC MDFs in this study. 

\section{DISCUSSION}

There is substantial evidence that GCs are the remnants of star formation events in galaxies, and are linked to the star formation, chemical enrichment, and merging histories of their parent galaxies \cite[e.g.,][]{mclaughlin99}. However, ever since {\it direct} photometry of spatially-resolved constituent stars in a dozen nearby galaxies became possible thanks to the {\it HST} and large ground-based telescopes, the discrepancy between the MDFs of GCs and field stars has remained a conundrum. If GCs mirror field stars across the galaxy histories, the MDFs of GCs and field stars should be similar. Thus, the curious disagreement in metallicity has been interpreted in the context of highly decoupled formation and evolution histories between GCs and constituent stars of their parent galaxies.

Current observational data and modeling point convincingly to nonlinear CMRs, which have significant implications for the interpretation of GC color distributions. We find that the strongly peaked [Fe/H] distributions inferred from 
nonlinear CMRs are qualitatively similar to the MDFs of field stars in the spheroidal component of nearby galaxies and to those produced by chemical evolution models of galaxies. However, whether the inferred GC MDFs represent the intrinsic, true ones is still unproven, and so it may be partly a coincidence. Nevertheless, if the MDFs obtained using stellar population models more closely represent the true GC MDFs, then this would change much of the current thought on the formation of GC systems and their host galaxies. The next two sections discuss what constraints would our findings pose on formation of GC systems (\S\S\,5.1) and their host galaxies (\S\S\,5.2). In Section 5.3, we present our view on the formation and evolution of GC systems and their parent galaxies. 

\subsection{What Do the Inferred GC MDFs Imply?}

Remarkable progress has occurred over the past few decades in our
understanding of extragalactic GC systems. 
One of the most important discoveries is that 
many galaxies show bimodality in their color distributions, 
leading to the notion that galaxies possess two distinct subpopulations of GCs. 

We showed, however, that the typical GC MDF shape derived from color distributions is unimodal and characterized by a sharp peak with a metal-poor tail. If confirmed, the inferred GC MDFs may appreciably reduce the demand for the separate formation mechanisms to explain the metal-poor and metal-rich division of GCs. 
We warn, though, that the sample used in this study is the GC systems in the Virgo galaxy cluster obtained from the {\it HST} ACS/WFC, WFPC2, and WFC3 observations. The field of view of ACS/WFC, for example, covers galaxies' haloes within R $\simeq$ 0.6, 0.8, 7.2, 8.3, 8.9, 9.4, and 10.6 $R_e$ (in $z$-band, Ferrarese \etal\ 2006) for the seven host luminosity bins (from the brightest bin to the faintest) of the ACSVCS, respectively. Nevertheless, the dearth of metal-poor GCs in MDFs for inner spheroids of giant ($M_B$ $\lesssim$ $-20$) elliptical galaxies and nearly the entire spheroids of normal ($-20$ $\lesssim$ $M_B$ $\lesssim$ $-15$) ellipticals suggest that the inheritance of metal-poor GCs via dissipationless accretion from dwarf satellites seems to be less significant than previously thought.

On the outskirts of giant galaxies, accretion of metal-poor GCs from low-mass satellites and/or from surrounding regions 
may be an important channel for a galaxy to add GCs on the metal-poor part of GC MDFs \citep{forte82,cote98,cote02,masters10,mglee10a}. 
Recall, however, that one of the main consequences of the nonlinear metallicity-color relations is that their steepness at the metal-poor end naturally creates a blue peak of GCs in color space, which is a direct cause of the conventional subpopulation of blue GCs.
Therefore, even for the outskirts of giant galaxies in cluster environments, whether or not accretion of metal-poor GCs is solely responsible for blue peaks of color distributions is still an open question.
The strongest evidence against accretion models is that the blue peak colors are correlated tightly to the host galaxy luminosity (Larsen \etal\ 2001; Strader, Brodie \& Forbes 2004; Peng \etal\ 2006).
The metal-poor relation implies that metal-poor GCs, although they formed at very high redshift and were accreted later on, already ``knew'' which galaxy they would ultimately belong to, and thus weakens the accretion scenario for the color bimodality.
Alternatively, the nonlinear metallicity-color relations scenario (Paper I) gives cohesive explanations for the observations that the mean colors of {\it both} blue and red GCs increase progressively for more luminous host galaxies.

Further counter-evidence of the accretion model is the significant fraction of blue GCs in massive cluster galaxies in relatively lower-density regions (Peng \etal\ 2006) and massive field galaxies in isolation (Cho \etal\ 2011). In such environments, galaxies have few neighboring lower-mass galaxies, and it would be difficult to acquire many metal-poor GCs via accretion. 
Therefore, the accretion process seems more important when it meets the three conditions: (a) the {\it outskirts} (rather than the inner, main bodies) of (b) {\it giant} (rather than dwarf) ellipticals orbited by a large number of low-mass satellites in (c) {\it cluster} (rather than isolated) environments. 
In this regard, wide-field, multiband studies of GC systems in cluster and field environments are clearly needed. Wide-field photometry of nearby {\it cluster} galaxies in CTIO 4-m $U$-band (H. Kim \etal\  2011, in prep.) and Subaru/MOIRCS {\it NIR} (S. Kim \etal\ 2011, in prep.), and a study of massive {\it field} galaxies in {\it HST}/ACS $g$ and $z$ (Cho et al. 2011) are done or in progress
to further investigate our alternative scenario.

Current hierarchical models of galaxy formation in the $\Lambda$CMD cosmology predict that several thousands of small building blocks were involved for the emergence of one single massive galaxy. This is significant because the extent of complexity may leave little room for the existence of just two GC subpopulations in each massive galaxy. Indeed, the unimodal, skewed MDFs arise naturally in an aggregate of a large number of protogalactic gas clouds from its virtually continuous chemical evolution through many successive rounds of star formation. The strongly peaked unimodal GC MDFs point to GC formation with a relatively short, quasi-monolithic timescale. Remarkably, the typical GC MDF shape emerges across three orders of magnitude in host galaxy mass. This suggests that the processes of GC formation and chemical enrichment are quite universal among a variety of GC systems.

\subsection{Do GC Systems Trace Star Formation in Galaxies?}

We also address the important issue of whether or not the formation of GCs is coupled with the bulk formation of the stellar population of host galaxies. Or, equivalently, does GC formation really mirror star formation in a galaxy? We have shown that the inferred GC MDFs agree reasonably well with the stellar MDFs of nearby galaxies and the MDFs produced by models of galaxy chemical evolution. The results suggest that the evolutionary histories of GC systems and their parent galaxies are strongly coupled, and thus share a more common origin and closer subsequent evolution than previously thought. 

An important aspect of the GC-host galaxy co-evolution issue concerns the GC-to-star offset, in the sense that GCs are on average more metal-poor than the unresolved light of an elliptical galaxy at the same radial location \cite[e.g.,][]{forte81,strom81,jordan04,tamura06,mglee08,forte09}. In light of the typical shape of GC MDFs proposed in this study, the lower value of the mean metallicity of GCs compared to that of field stars should not be attributed to the number excess of the metal-poor GCs. Alternatively it is more likely that, at a given radial location, the GC MDF on the whole is shifted toward the metal-poor side with a peak at a lower [Fe/H] value, compared to the field-star MDF. 

The GC-to-star offset would point to a picture in which GCs are the remnants of vigorous star burst events in the early stages of galaxy formation, and thus preferentially trace the {\it major} mode of star formation in galaxies. If so, GC formation was less prolonged than field star formation. Recent observations show that GCs were at least an order of magnitude more massive at birth than now \cite[e.g.,][]{conroy11}, and the large masses may have been the cause of the earlier truncation of their formation process than that of stars. Consequently, the chemical enrichment process of a GC system appears to have ceased somewhat earlier than that of the field stellar population in each star formation episode. 

We further showed a possible age difference among GC systems, in that the GC systems in fainter galaxies are on average younger. We refer to this as ``GC system downsizing.''  The GC system downsizing phenomenon appears to further support the similar nature shared by stellar populations of GCs and field stars. 
Interestingly, recent observations reveal that the GC systems in Milky Way satellites and some Milky Way GCs believed to be accreted from satellite dwarf galaxies are relatively younger than the majority of Galactic GCs (e.g., Mar{\'{\i}}n-Franch \etal\ 2009)

Galaxy downsizing is generally defined by more {\it prolonged} residual star formation in fainter galaxies.
However, the ``GC system downsizing'' involves the idea that the first generation of GCs in fainter hosts are created later than the counterpart in brighter galaxies, and may provide further information on galaxy downsizing itself.  
Provided that GC systems became attached to the present host galaxies from the beginning, we can speculate that not only the first GCs but also {\it first field stars} were formed earlier in brighter galaxies than fainter galaxies. 
That is, massive galaxies today were likely where conditions first favored star formation, thus suggesting a prolonged epoch of galaxy formation in the universe. 
Combined with the fact that GCs and field stars in brighter galaxies have higher metallicities, a picture emerges in which the formation and accompanying metal enrichment of both GCs and halo stars seem to have started earlier and proceeded more rapidly and efficiently in massive galaxies, presumably in denser environments. 

It is also important to note that some galaxies have substructure in their stellar MDFs and are thus more complex than a smooth transition from a metal-rich dominance to metal-poor with increasing radius. Examples include the Milky Way \citep{ibata01,majewski03,yanny03,ivezic08} and M31 \citep{kalirai06,koch08,mcconnachie09}. The stellar MDF of the elliptical galaxy NGC 3379 (Figure 5$c$) also shows a fine substructure \citep{harris07b}. The remote outer haloes of these galaxies are inhomogeneous in terms of surface density and show different metallicity distributions from the inner haloes. This indicates that the stellar populations in the outer haloes of galaxies are not well mixed, and in turn supports the build up of halo formation via satellite accretion \citep{forte82,cote98,cote02,masters10}, which can have mixing time scales of a few Gyr in these outer regions \citep{johnston95}. 

The evidence of the satellite accretion indicates that the outskirts of galaxies do not all build up their stellar populations in one single way. For instance, the NGC 3379 observation fits into a model in which its outskirts were formed by a combination of earlier dissipative mergers and later accretion of dwarf satellites. The remote halo of NGC 3379 has been on the fringes of active, violent gas-rich mergers at the early epochs of galaxy formation, and later experienced dry accretion of stars and GCs from its metal-poor satellite dwarfs. Therefore, one should not assume that the stellar MDFs are unimodal at {\it all} positions, and our new interpretation would be more applicable to the inner, main bodies of galaxies than their remote outskirts. 

\subsection{An Alternative View on the Formation and Evolution of GC Systems and their Parent Galaxies}

Our results may be an important step forward in resolving the long-standing disagreement between GCs and field stars, 
and in reconstructing the history of GC systems and their parent galaxies.  
Although the sample used in this study (the {\it HST} ACS/WFC, WFPC2, and WFC3 photometry for the GC systems in the Virgo galaxy cluster) confines our discussion to $R$ $\lesssim$ $R_e$ for giant ellipticals and $\lesssim$ 10 $R_e$ for normal ellipticals, our findings suggest that GC systems and their parent galaxies have shared a more common origin than previously thought, and hence greatly simplify theories of galaxy formation. 
The star formation and accompanying chemical evolution were virtually continuous 
via aggregates of a large number of protoclouds 
and via repeated gas-rich mergers with other galaxies, leading to the unimodal, skewed MDFs of both stars and GCs.
The metal enrichment of both stars and GCs proceeded more rapidly and efficiently in massive galaxies,
resulting in more metal-rich stars and GCs in those galaxies. 
The observed radial metallicity gradients are understood if the chemical enrichment in the dense centers was more rapid and efficient than in the less-dense outskirts. 
The typical GC MDF shape emerges across three orders of magnitude in the host galaxy mass, suggesting that the processes of GC formation and chemical enrichment are quite universal among various GC systems,
at lease for the inner, main spheroids of giant ellipticals and the nearly entire haloes of normal ellipticals.

The histories of GCs and their host galaxies are reconstructed as follows.

\begin{enumerate}

\item
{\bf Giant ellipticals' inner, main haloes: }  
The inner, main spheroids (R $\lesssim$ 1 $R_{e}$) of today's giant elliptical galaxies ($M_B$ $\lesssim$ $-$20)
were first created in dense regions of the universe via dissipational mergers of a large number of protogalactic gas clouds (e.g., Searl \& Zinn 1978).
Their GC systems with the MDF peaks at [Fe/H] $\gtrsim$ $-0.7$ (Figures 3 and 4) were formed together with stars in the galaxies.
The GC systems from low to high metallicities (i.e., both blue and red GCs), as a whole, do not need two separate mechanisms for formation. 
There is strong similarity between blue and red GCs in their mass function; for the less evolved high-mass part of the mass function, these are approximately power laws with indices of  $-1.8$ to $-2$. Although power-law distributions are a consequence of a variety of physical processes, their nearly identical power indices may support an identical formation history of blue and red GCs. The chemical enrichment of the galaxies was rapid, and the difference in the formation epochs of the metal-poor and metal-rich ends of MDFs should be small ($\lesssim$ 1 Gyr), as evidenced by observations (e.g., Jord\'an \etal\ 2002). 
In this view, the metal-poor GCs in massive galaxies are the first generation of GCs in the universe.

\item 
{\bf Normal ellipticals' entire haloes: }  
The spheroids (out to $R\lesssim$ 10 $R_e$) of normal galaxies ($M_B$ $\gtrsim$ $-$20) were created later than those of massive ones via self-collapses or mergers of protogalactic clouds that did not take part in massive galaxy formation early on. They were outside of tumultuous, dense regions, and have evolved rather independently from central massive galaxies.  
Their GC systems with the MDF peaks at [Fe/H] $\lesssim$ $-0.7$ (Figures 3 and 4) were formed together with stars in the low-mass galaxies.
As a result, the first generation of both stars and GCs in low-mass galaxies are younger than those of massive galaxies, 
indicating a prolonged epoch of galaxy formation in the universe. 
This picture is supported by recent observational evidence that the GC systems in Galactic satellites and some Galatic GCs believed to be accreted from satellite dwarf galaxies are relatively younger than the majority of Milky Way GCs (e.g., Mar{\'{\i}}n-Franch \etal\ 2009).

\item 
{\bf Remote outskirts of giant ellipticals in dense environments: }  
The remote outer haloes of giant elliptical galaxies have been on the fringes of active, violent gas-rich mergers at the early epochs of galaxy formation, and later experienced dissipationless accretion of stars and GCs from its metal-poor satellite galaxies \citep{forte82,cote98,cote02,masters10,mglee10a}
and/or from the surrounding regions (Tamura \etal\ 2006a,b; Bergond \etal\ 2007; Schuberth \etal\ 2008; Lee \etal\ 2010b; West \etal\ 2011).
Such GCs with an accretion origin tend to be of lower metallicity and on highly elongated orbits with high energy. 
The GCs generally favor the extended outskirts of giant galaxies (Lee et al. 1998, 2008b; Lee 2003; Dirsch et al. 2005), 
although they, on highly elongated orbits, must penetrate into the inner body of galaxies.
This explains the observations that the bimodality in GC colors corresponds to a change in the kinematic properties of the GCs, such that the blue GCs are dynamically hotter and there is a ``bimodal'' distribution of velocity dispersion. 
The demarcating radii between the inner, main part (the product of mergers) and the outskirts (the product of merger plus accretion) are not sharp and vary galaxy-to-galaxy depending on their individual histories of mergers and accretions.
This is in line with the considerable diversity in the kinematics of the GC systems in giant elliptical galaxies (e.g., Hwang \etal\ 2008; Lee \etal\ 2010a).


\end{enumerate}

Observationally, the intimately coupled histories of the GCs and constituent field stars of elliptical galaxies is encouraging in the context of mapping their metal contents, which is one of the most important yet least known properties of external galaxies. Even with the forthcoming larger telescopes, individual stars can be spatially resolved only in a few tens nearby galaxies \citep{tolstoy06}. Hence, we anticipate that the estimation of GC metallicities based on their colors, equipped with a proper correction for the GC-to-star offset, will become a preferred practical method in the quest to comprehend the evolution of galaxies beyond the nearby universe. Further refinement of the exact shape of the CMRs for GCs, and the resulting implications for the formation of GC systems and galaxies, should be the topic of much future work.

\clearpage
\acknowledgments 
We would like to thank the anonymous referee for useful comments and suggestions.
SJY acknowledges support 
from Mid-career Researcher Program (No. 2009-0080851) and Basic Science Research Program (No. 2009-0086824) through the National Research Foundation (NRF) of Korea grant 
funded by the Ministry of Education, Science and Technology (MEST), 
and from the Korea Astronomy and Space Science Institute Research Fund 2011. 
SJY and YWL acknowledge support by the NRF of Korea to the Center for Galaxy Evolution Research. 
JPB thanks the Yonsei University Center for Galaxy Evolution Research for the generous hospitality during his visit.
EWP acknowledges support from the Peking University 985 fund and grant 10873001 
from the National Natural Science Foundation of China. 
SJY would like to thank Daniel Fabricant, Charles Alcock, Jay Strader, Dongsoo Kim, Jaesub Hong for their 
hospitality during his stay at Harvard-Smithsonian Center for Astrophysics as a Visiting Professor in 2011--2012.


\clearpage
\begin{figure*}
\begin{center}
\includegraphics[width=17.7cm]{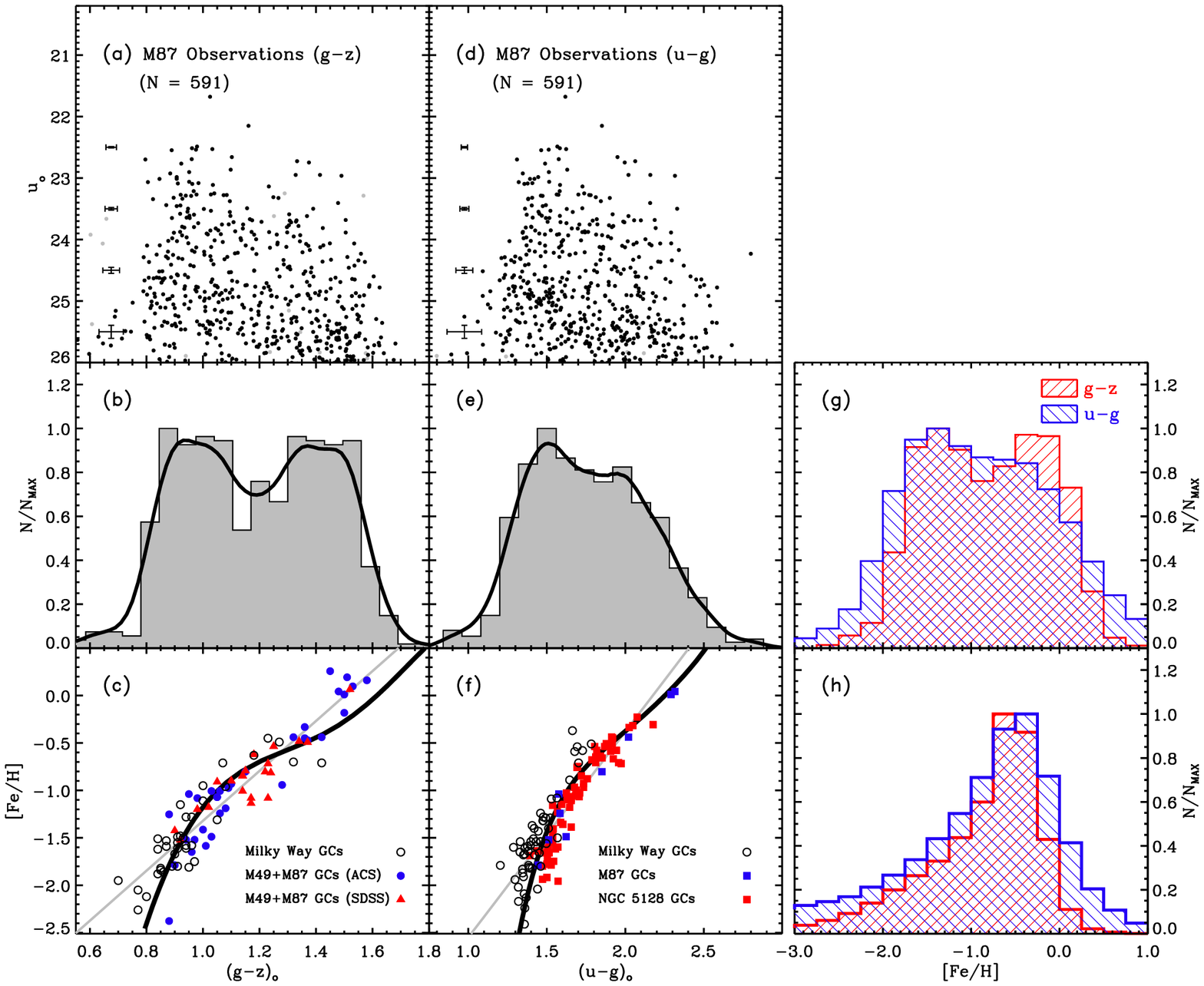}   
\end{center}
\end{figure*}

\clearpage
\begin{figure*}
\begin{center}
\caption{ \small
The color-magnitude diagrams, and color and inferred metallicity distributions for the GC system in giant elliptical galaxy, M87.
($a$) The $g-z$ vs. $u$ diagram for the GCs in M87. The {\it HST}/WFPC2 archival F336W ($u$-band) images were used to derive $u$ (ABMAG) for GC candidates. The $u$-band catalog was matched with ACS/WFC $g$- and $z$-band (ABMAG) photometry of Jord\'{a}n et al (2009). The error bars represent the observational uncertainties of the corresponding magnitude bins. Jord\'{a}n et al (2009) selected {\it bona-fide} GCs with their magnitudes, $g-z$ colors, and sizes. We further employed color cuts in the $u$-band colors to filter out contaminating sources, especially background star-forming galaxies. The 591 GCs that have reliable $u$, $g$, and $z$ measurements in common are only used, and the sample is $u$-band limited ($\sigma$$_u$ $<$ 0.2).
($b$) The {\it HST}/ACS $g-z$ color distribution for the M87 GCs. Distribution is normalized to the maximum value. Solid line is a smoothed histogram with a Gaussian kernel with $\sigma$($g-z$) = 0.05. 
($c$) The relationship between $g-z$ and [Fe/H] for the 40 low-extinction Galactic GCs (open circles), 33 M49 and M87 GCs with ACSVCS photometry (blue circles), and 22 M49 and M87 GCs with SDSS photometry (red triangles). The references to the observed data used in the relation are summarized in Table 1. The 5th-order polynomial fits to our model prediction for 13.9-Gyr GCs is overlaid (thick solid line). The $\alpha$-element enhancement parameter, [$\alpha$/Fe], is assumed to be 0.3. Thin grey straight line is for the linear least-squares fit to the data. 
($d$, $e$, and $f$) The same as ($a$--$c$), but for the $u-g$ colors. In ($e$), solid line is a smoothed histogram with a Gaussian kernel with $\sigma$($u-g$) = 0.15. In ($f$), open circles, blue and red filled squares represent GCs in the Milky Way, M87, and NGC 5128, respectively. The $u-g$ colors of the GCs in the Milky Way and NGC 5128 were obtained from their $U-B$ colors via the equation, ($u-g$) = 1.014 ($U-B$) + 1.372, derived from model data for synthetic GCs with combinations of age (10 $\sim$ 15 Gyr of 1 Gyr intervals) and [Fe/H] ($-2.5$ $\sim$ 0.5 dex of 0.1 dex intervals). 
($g$) The inferred GC MDFs based on the simple linear fit to the CMRs shown in ($c$) and ($f$). The red and blue hashed histograms are obtained using the $g-z$ distribution in ($b$) and the $u-g$ distribution in ($e$), respectively. Distributions are normalized to the maximum values. 
($h$) The same as ($g$), but with the inferred GC MDFs using the nonlinear color-to-metallicity transformations predicted by the YEPS models in ($c$) for the $g-z$ distribution and in ($f$) for the $u-g$ distribution.\label{fig1}}
\end{center}
\end{figure*}

\clearpage
\begin{figure*}
\begin{center}
\includegraphics[width=17.7cm]{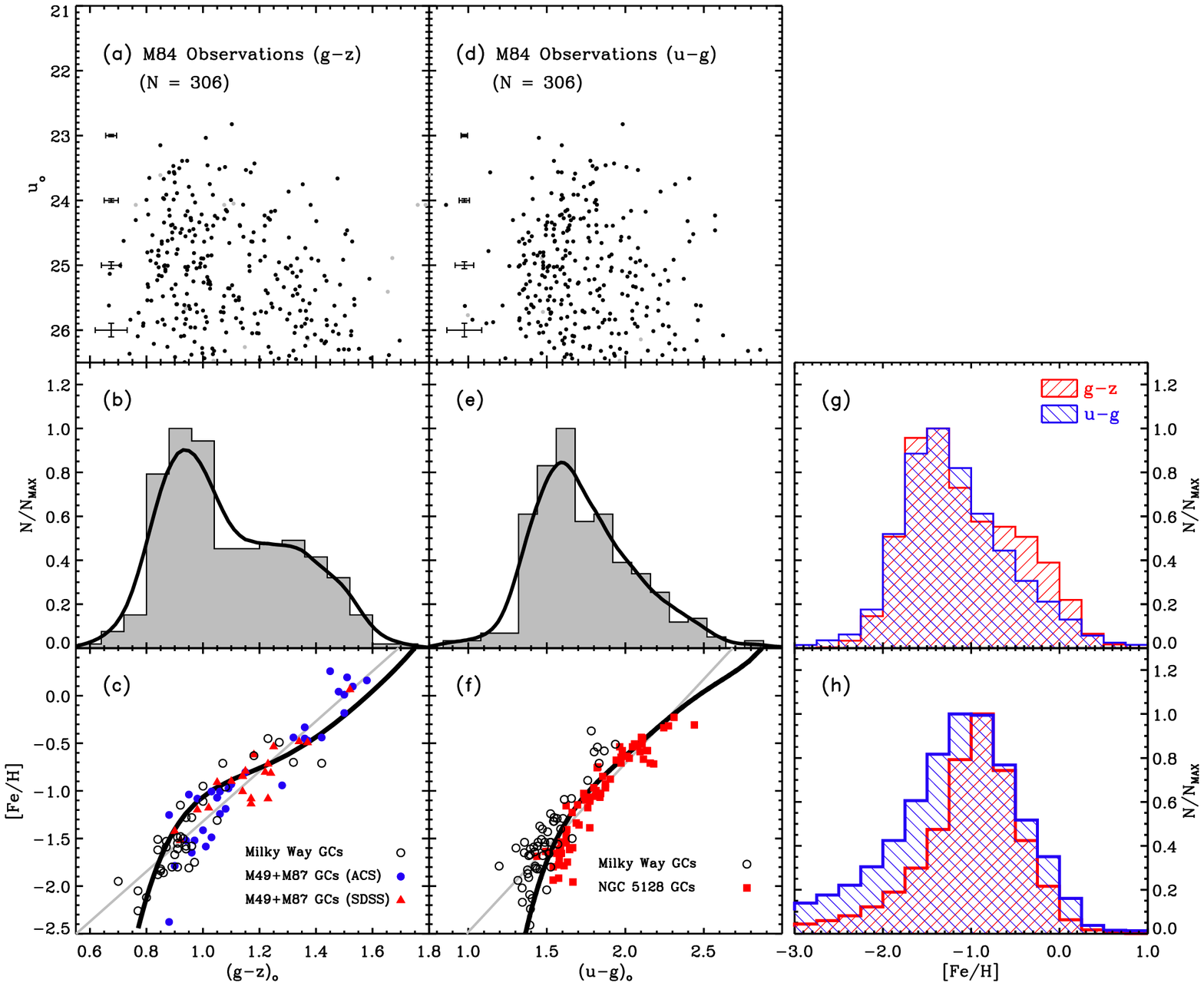}
\end{center}
\end{figure*}

\clearpage
\begin{figure*}
\begin{center}
\caption{ 
The same as Figure 1, but for the M84 GCs. 
Their $u$ (ABMAG) were obtained from {\it HST}/WFC3 archival images. 
The 306 GCs that have reliable $u$, $g$, and $z$ measurements in common are only used, and the sample is $u$-band limited ($\sigma$$_u$ $<$ 0.2).
In ($b$) and ($e$), the 5th-order polynomial fits to our model prediction for 13-Gyr GCs is overlaid (thick solid line).
Note that, in ($f$) the modeled $u-g$ CMR differs from that in Figure 1$f$ due to the difference in filter throughput between F336W's on WFPC2 and WFC. 
The main difference lies in that the WFC F336W has much less red leak than the WFPC2 F336W. 
The $u-g$ colors of the GCs in the Milky Way and NGC 5128 were converted from their $U-B$ colors via the equation, ($u-g$) = 1.296 ($U-B$) + 1.412, derived from model data for synthetic GCs with combinations of age (10 $\sim$ 15 Gyr of 1 Gyr intervals) and [Fe/H] ($-2.5$ $\sim$ 0.5 dex of 0.1 dex intervals).
The references to the observed data used in the relation are summarized in Table 1.
\label{fig2}}
\end{center}
\end{figure*}

\clearpage
\begin{figure*}
\begin{center}
\includegraphics[width=11cm, angle=180]{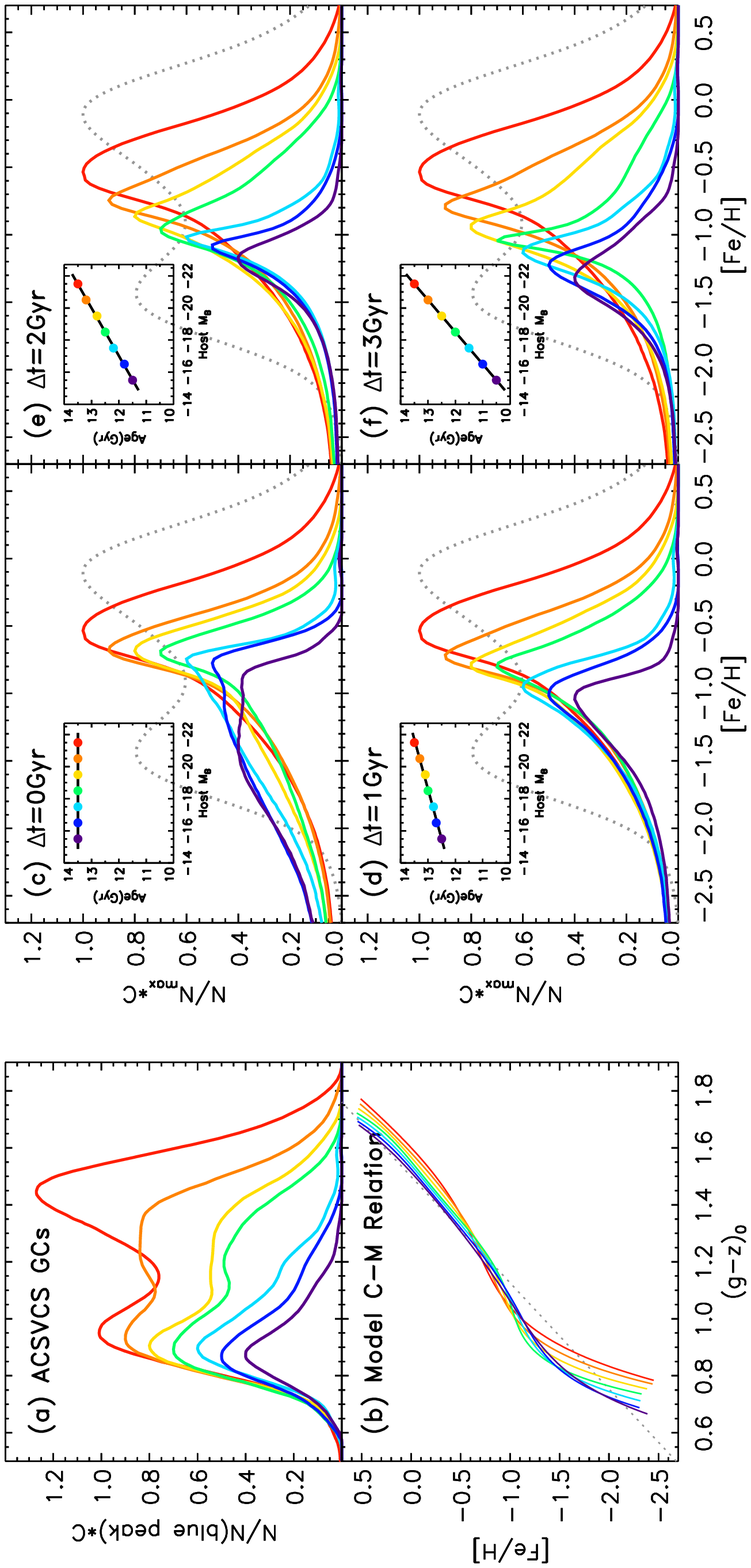}
\end{center}
\end{figure*}

\clearpage
\begin{figure*}
\begin{center}
\caption{ 
Color and metallicity distributions of GCs in ACSVCS galaxies.
($a$) The observed color histograms of GC systems for seven bins of host galaxy magnitude. The data are the same as in Figure 6 of \citet{peng06} and now listed in Table 4. The histograms are normalized by the GC number at their blue peaks, and multiplied by constants, $C$, for clarity. The magnitude bins are 1 mag wide and extend from $M_B$ $\simeq$ $-$21.5 ($-$22 $\leq$ $M_B$ $<$ $-$21, red, $C$ = 1.0) to $\simeq$ $-$15.5 ($-$16 $\leq$ $M_B$ $<$ $-$15, purple, $C$ = 0.4). A Gaussian kernel of $\sigma$($g-z$) = 0.05 is applied.
($b$) Similar to Figures 1$c$ and 2$c$, but the model predictions for various ages ranging from 10.5 Gyr (purple) to 13.5 (red) by equal age intervals of 0.5 Gyr (see Table 2). The $\alpha$-element enhancement parameter, [$\alpha$/Fe], is assumed to be 0.3. The straight dotted line, again, is for the linear fit to the observational data points shown in Figures 1$c$ and 2$c$, which are not plotted here for clarity.
($c$) The inferred MDFs of GCs in the same mag bins as in ($a$), multiplied by corresponding constants, $C$. 
Each GC MDF is obtained from the corresponding color distribution using the $g-z$ to [Fe/H] conversion,
and the same color code as in ($a$) is applied.
The modeled ages for the host mag bins are given in the insets. In ($c$), the age is assume to be constant at 13 Gyr regardless of host luminosities between $M_B$ $\simeq$ $-$15.5 (purple) and $-$21.5 (red). The grey dotted histogram represents the inferred MDF for the brightest (i.e., $M_B$ $\simeq$ $-$21.5) bin based on the simple straight fit in ($b$), and is supposed to be compared to the red solid histogram. 
($d$, $e$, $f$) The same as ($c$), but with age differences of 1.0, 2.0, and 3.0 Gyr, respectively, between the faintest ($M_B$ $\simeq$ $-$15.5, purple) and brightest ($-$21.5, red) bins. The brightest ($M_B$ $\simeq$ $-$21.5, red) bin is set to be 13.5 Gyr. The data for the histograms in ($d$) are given in Table 5.
\label{fig3}}
\end{center}
\end{figure*}

\clearpage
\begin{figure*}
\begin{center}
\includegraphics[width=11cm, angle=180]{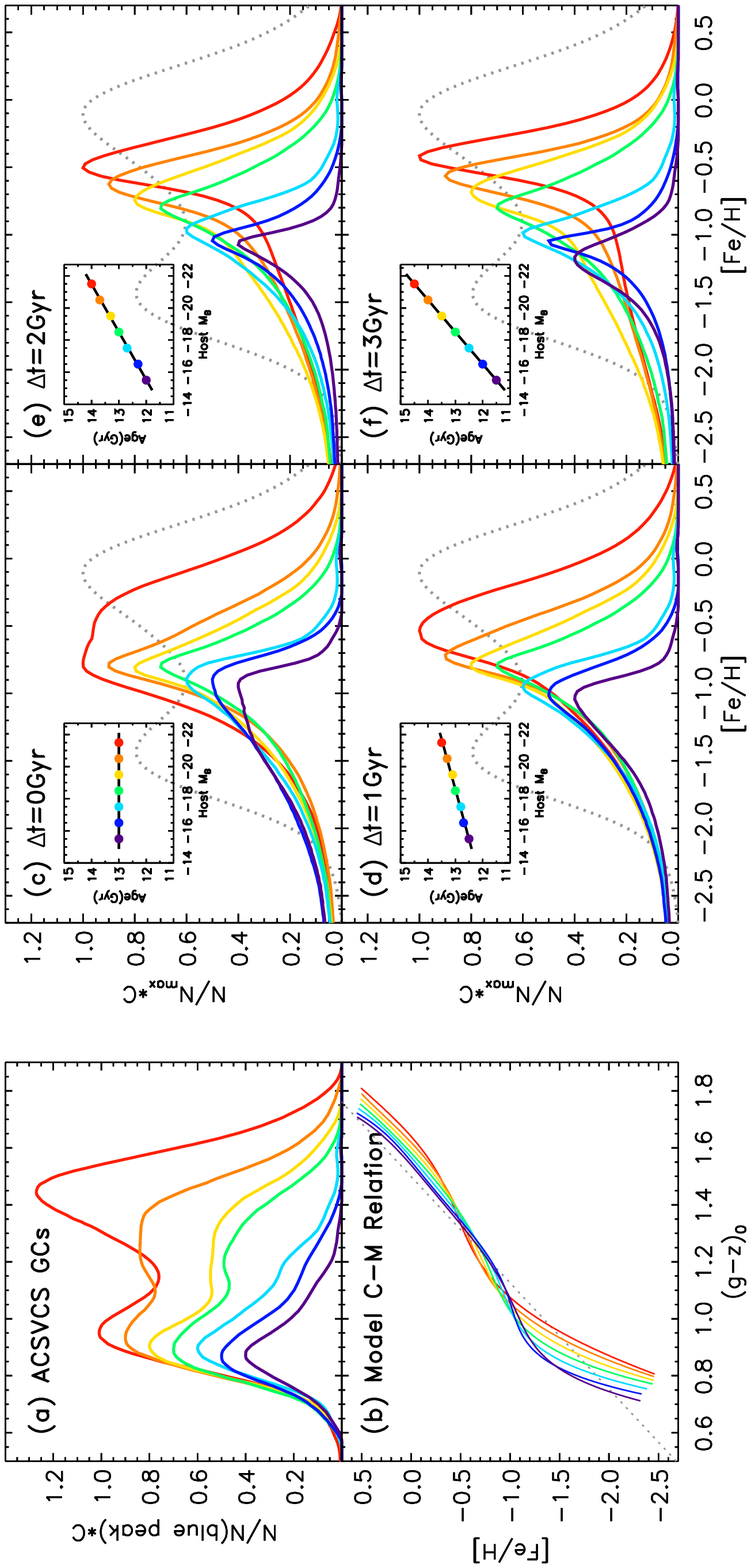}
\end{center}
\end{figure*}

\clearpage
\begin{figure*}
\begin{center}
\caption{ 
The same as Figure 3, but under a different assumption on the host age sequence from the brightest bin to faintest. 
It is assumed that the center bin, i.e., the fourth brightest ($M_B$ $\simeq$ $-$18.5, green) bin, is set to be 13 Gyr.
In ($b$), the YEPS model CMRs are displayed with ages ranging from 11.5 Gyr (purple) to 14.5 (red) by equal age intervals of 0.5 Gyr.
The data for the histograms in ($d$), which is identical to Figure 3$d$, are listed in Table 5.
\label{fig4}}
\end{center}
\end{figure*}

\clearpage
\begin{figure*}
\begin{center}
\includegraphics[width=17.2cm]{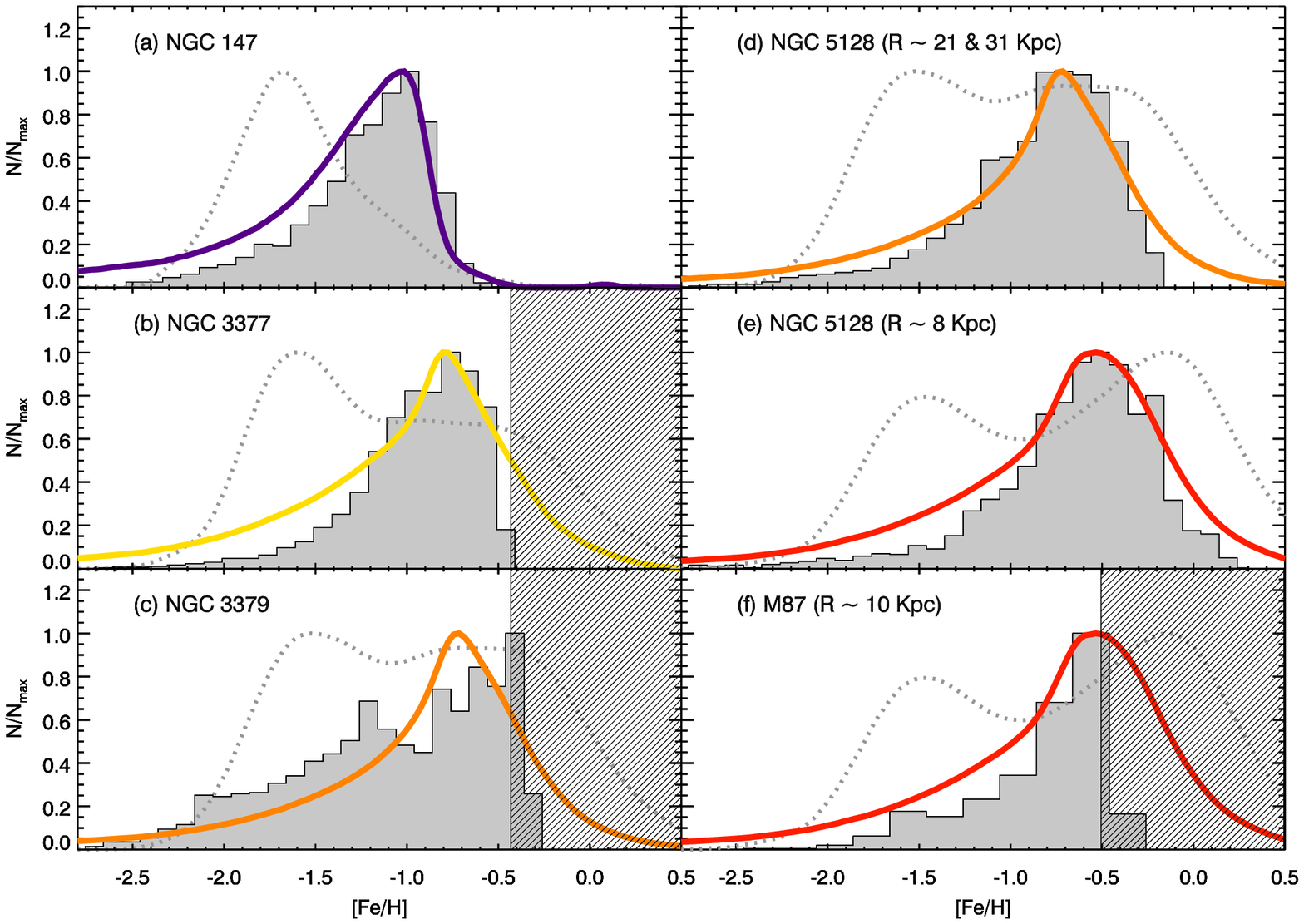}
\end{center}
\end{figure*}

\clearpage
\begin{figure*}
\begin{center}
\caption{ 
Comparison between MDFs of GCs and field stars. The inferred MDFs of ACSVCS GCs (thick curves) are taken from Figure 3$d$ (see also Figure 4$d$ and Table 5), whereas those of field stars (grey histograms) are obtained from photometric studies on resolved red-giant stars in individual nearby galaxies. For a comparison, the dotted curve in each panel represents the MDF of ACSVCS GCs for the corresponding host luminosity bin that is derived based on the simple, straight fit shown in Figures 1$c$ and 2$c$. The field star MDFs are measured in terms of [m/H] $\equiv$ Log(Z/Z$_\sun$) -- Log(X/X$_\sun$), from which [Fe/H] is obtained using the equation, [Fe/H] =  Log(Z/Z$_\sun$) -- Log(X/X$_\sun$) -- 0.723 [$\alpha$/Fe] = [m/H] -- 0.217, for [$\alpha$/Fe] = 0.3 (Table 2 of \citet{yckim02}, and see also \citet{shetrone01}). The stellar metallicity is in the Zinn-West scale \citep{zinn84}. All distributions are normalized to the maximum values. 
($a$) The stellar MDF of NGC 147 \citep{harris07a,harris07b} (a Local Group dwarf elliptical, $M_B$ = $-$14.8) is compared to that of ACSVCS GCs in host galaxies with $M_B$ $\simeq$ $-$15.5 (purple). 
($b$) The same as ($a$), but for host galaxies in $M_B$ $\simeq$ $-$19.5 bin (yellow) and for field stars in NGC 3377 \citep{harris07a,harris07b} (a Leo group elliptical, $M_B$ = $-$19.2). The hashed area at [Fe/H] $>$ $-$0.4 marks the region of low credibility on the stellar metallicity measurement due to photometric incompleteness.
($c$) The same as ($b$), but for host galaxies in $M_B$ $\simeq$ $-$20.5 bin (orange) and for field stars in NGC 3379 \citep{harris07a,harris07b} (a Leo group elliptical, $M_B$ = $-$20.6).
($d$) The same as ($a$), but for host galaxies in $M_B$ $\simeq$ $-$20.5 bin (orange) and for field stars in the halo of NGC 5128 \citep{rejkuba05}, a nearby field giant elliptical galaxy, with $M_B$ = $-$20.9. The observed regions are $\sim$ 21 and $\sim$ 31 Kpc away from the galaxy center.
($e$) The same as ($d$), but for host galaxies in $M_B$ $\simeq$ $-$21.5 bin (red) and for field stars in the halo of, again, NGC 5128 \citep{rejkuba05}. The observed region is $\sim$ 8 Kpc away from the galaxy center.
($f$) The same as ($e$), but for host galaxies in $M_B$ $\simeq$ $-$21.5 bin (red) and for field stars in the halo of M87 \citep{bird10} with $M_B$ = $-$21.4. The observed regions are $\sim$ 10 Kpc away from the galaxy center. 
The hashed area at [Fe/H] $>$ $-$0.5 marks the region of low credibility on the stellar metallicity measurement due to photometric incompleteness.
\label{fig5}}
\end{center}
\end{figure*}

\clearpage
\begin{figure*}
\begin{center}
\includegraphics[width=17cm]{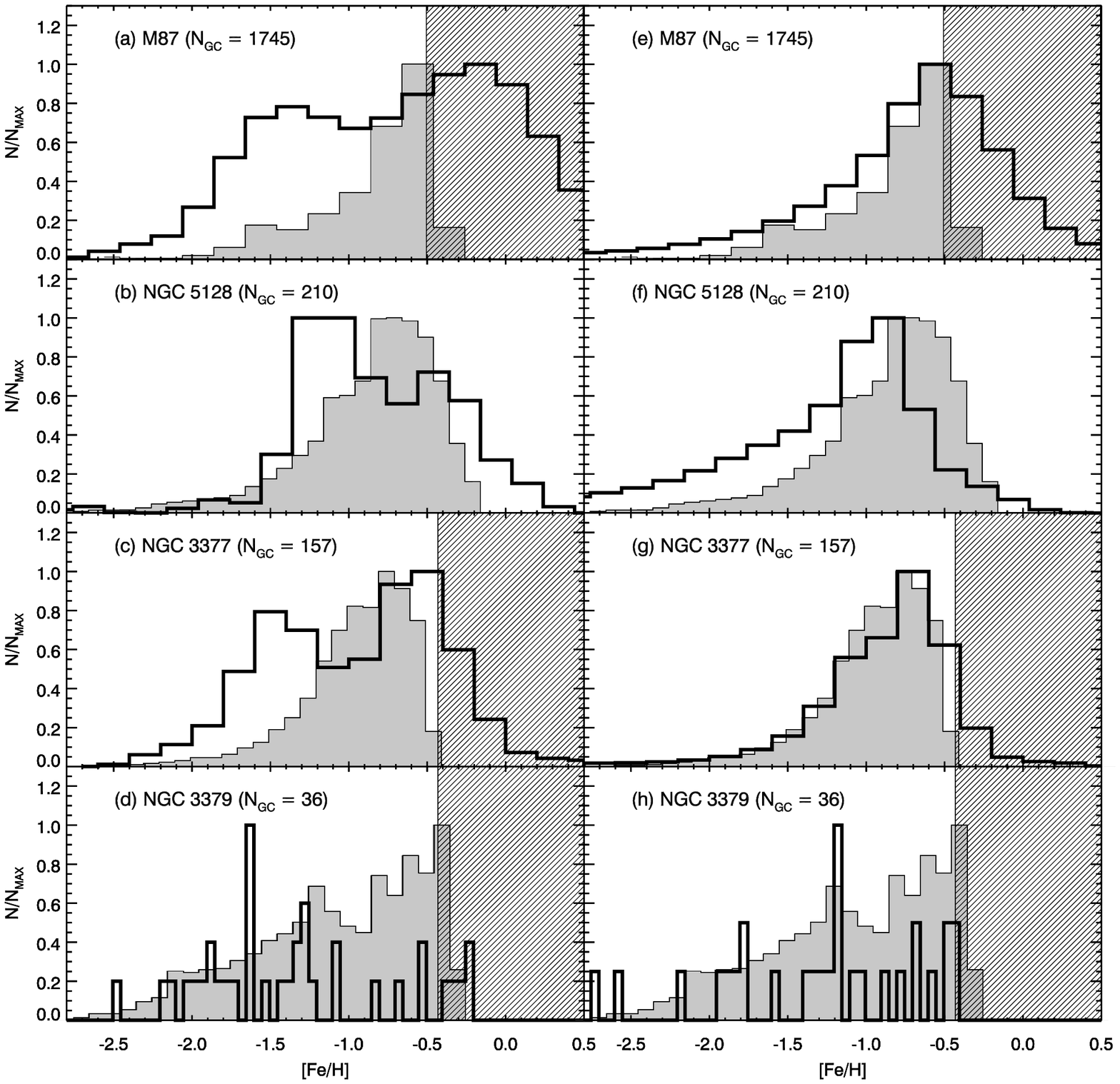}
\end{center}
\end{figure*}

\clearpage
\begin{figure*}
\begin{center}
\caption{ 
Comparison between the MDFs of field stars of galaxies and their own GC systems for M87 (the top row), NGC 5128 (the second row), NGC 3377 (the third row), and NGC 3379 (the bottom row). Among the galaxies in Figure 5, the four are those for which both the GC MDF and stellar MDF are currently available.  The stellar MDFs (grey histograms) are identical to those in Figures 5$b$, 5$c$, 5$d$, and 5$f$. The left-hand panels present the inferred GC MDFs based on the traditional linear color-to-metallicity transformations. By contrast, the right-hand panels show the inferred GC MDFs based on the inflected relations from the YEPS model. All distributions are normalized to the maximum values.
($a$) The stellar MDF (grey histogram) of M87 \citep{bird10} is compared to the MDF of 1745 GCs in the same galaxy (open thick histogram) obtained using the simple linear [Fe/H] vs. $g-z$ relation in Figure 1$c$. The stars were sampled in an inner region (R $\simeq$ 10 Kpc) and the GCs were in the inner halo ($\lesssim$ 10 Kpc). 
($b$) The stellar MDF (grey histogram) of NGC 5128 \citep{rejkuba05} is compared to the GC MDF (open thick histogram) obtained from CTIO Blanco 4-m $B-I$ histogram for 210 GCs in the same galaxy \citep{peng04a,peng04b}, using an empirical linear relation, [Fe/H] = 1.88 ($B-I$)$_0$ -- 4.11 \citep{spitler08}. The stars were sampled in an inner region (R $\simeq$  20 and 30 Kpc) and the GCs were in the entire halo of this galaxy.
($c$) The stellar MDF (grey histogram) of NGC 3377 \citep{harris07a,harris07b} is compared to the GC MDF (open thick histogram) obtained from {\it HST}/ACS $g-z$ histogram for 157 GCs in the same galaxy \citep{cho11} using the simple linear [Fe/H] vs. $g-z$ relation in Figure 1$c$. 
($d$) The stellar MDF (grey histogram) of NGC 3379 \citep{harris07a,harris07b} is compared to the GC MDF (open thick histogram) obtained from Rhode \& Zepf (2004)'s $B-R$ histogram for 36 GCs with an improved empirical relation, [Fe/H] = 3.13 ($B-R$)$_0$ -- 5.04 \citep{harris07b}. 
($e$, $f$, $g$, and $h$) The same as ($a$--$d$), but for the GC MDFs of M87, NGC 5128, NGC 3377, and NGC 3379 are inferred from the YEPS model CMRs for $g-z$ (13.5 Gyr), $B-I$ (12.5 Gyr), $g-z$ (12.8 Gyr), and $B-R$ (12.5 Gyr), respectively. 
\label{fig6}}
\end{center}
\end{figure*}

\clearpage
\begin{figure*}
\begin{center}
\includegraphics[width=14cm]{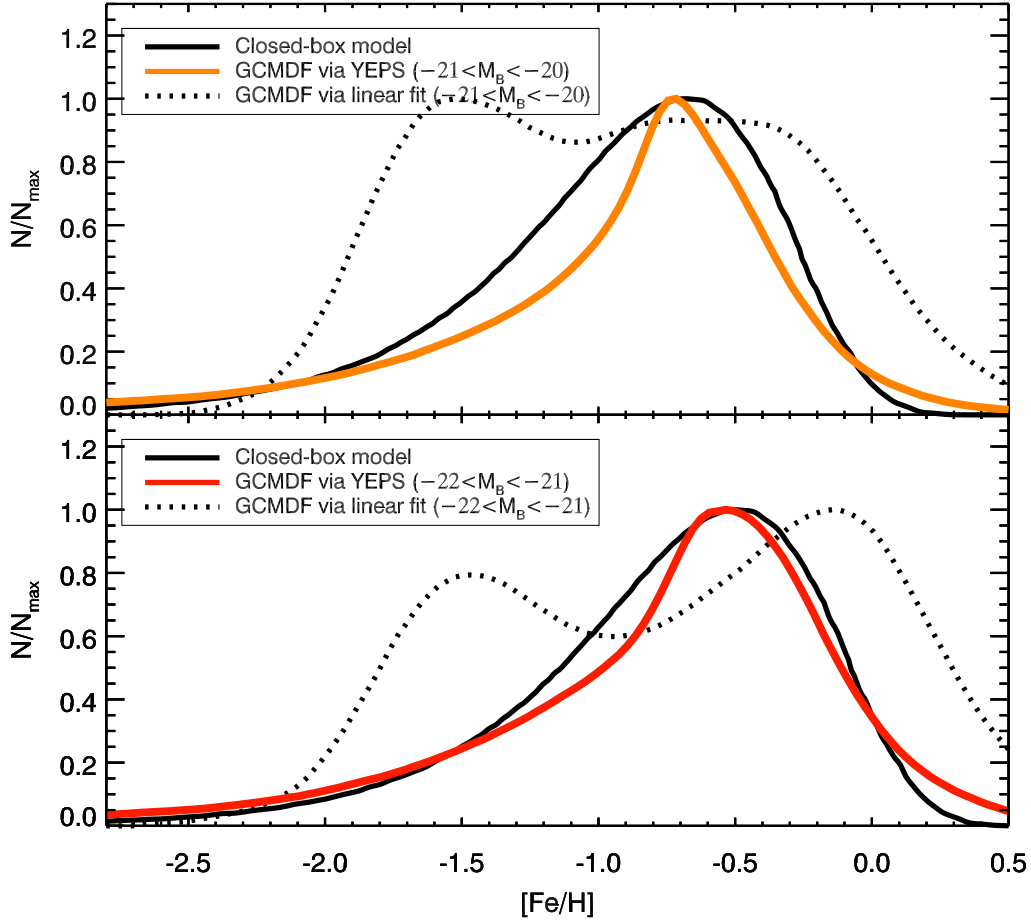}
\caption{
Comparison between inferred GC MDFs and those from chemical enrichment models. The MDFs of ACSVCS GCs are taken from Figure 3$d$ (see also Figure 4$d$ and Table 5). All distributions are normalized to the maximum values. 
($Upper$) The MDF of ACSVCS GCs in the second brightest host bin with $M_B$ $\simeq$ $-$20.5 (orange) is compared to that of a simple closed-box model with yield [Fe/H] = $-$0.69 (black solid line). For a comparison, dotted histogram represents the inferred MDF for the same $M_B$ $\simeq$ $-$20.5 bin based on the simple straight fit shown in Figures 1$c$ and 2$c$.
($Lower$) The same as the upper panel, but for the brightest host bin with $M_B$ $\simeq$ $-$21.5 (red) and for model with yield [Fe/H] = $-$0.51 (black solid line). For a comparison, dotted histogram represents the inferred MDF for the same $M_B$ $\simeq$ $-$21.5 bin based on the simple straight fit shown in Figures 1$c$ and 2$c$.
\label{fig7}}
\end{center}
\end{figure*}


\clearpage
\begin{table}
\footnotesize
\begin{center}
\caption{References to the observational data for the [Fe/H] vs. color relations shown in Figures 1 and 2.\label{tbl-1}}
\vspace{0.5cm}
\begin{tabular}{lllll}
\tableline
\tableline
{Relations}  &  Galaxy Name & &\multicolumn{2}{l}{References and Selection Criteria} \\
\tableline
			      		 		&   				& & Spectroscopic [Fe/H] 			& Broadband Color 							\\
\tableline
Figures 1$c$ \& 2$c$			& Milky Way         &	& 1, 2 						& 1, 2 									\\
The [Fe/H] vs. $g-z$ relation		& M49	 	&	& 1, 2 						& 1, 2 									\\
							& M87 			& & 1, 2 						& 1, 2 									\\
\tableline
Figure 1$f$ 					& Milky Way 		& & 3 							& 3 [($U-B$)\tablenotemark{a}, E($B-V$) $<$ 0.3]	\\
The [Fe/H] vs. $u-g$ relation		& NGC 5128		& & 4, 5 [$t$ $>$ 8 Gyr, S/N $>$ 10]	& 6 [($U-B$)\tablenotemark{a}]					\\
($u$ = HST/WFPC2 $F336W$)		& M87  		&	& 1, 2 						& This study [($u-g$)] 						\\
\tableline
Figure 2$f$ 					& Milky Way 	&	& 3 							& 3 [($U-B$)\tablenotemark{b}, E($B-V$) $<$ 0.3]	\\
The [Fe/H] vs. $u-g$ relation		& NGC 5128	&	& 4, 5 [$t$ $>$ 8 Gyr, S/N $>$ 10]	& 6 [($U-B$)\tablenotemark{b}]					\\
($u$ = HST/WFC3 $F336W$)		& M87  		&	& 1, 2 						& This study [($u-g$)] 						\\
\tableline
\tablenotetext{a}{The equation, ($u-g$) = 1.014 ($U-B$) + 1.372, is used for {\it HST}/WFPC2 $u$-band F336W (see Figure 1).}
\tablenotetext{b}{The equation, ($u-g$) = 1.296 ($U-B$) + 1.412, is used for {\it HST}/WFC3 $u$-band F336W (see Figure 2).}
\end{tabular}
\tablerefs{(1) \citet{peng06}; (2) Paper I; (3) Harris \etal\ (1996, the 2010 edition); (4) \citet{beasley08}; (5) \citet{chung11}; (6) \citet{peng04a,peng04b}.}
\end{center}
\end{table}

\clearpage
\begin{table}
\scriptsize
\begin{center}
\caption{The $g-z$ color from the YEPS model for GCs with ages ($t$) ranging from 9 Gyr to 14 Gyr by steps of 0.5 Gyr.\label{tbl-2}}
\vspace{0.5cm}
\begin{tabular}{rrrrrrrrrrrrrr}
\tableline
\tableline
{[Fe/H]}  &  \multicolumn{11}{c}{$g-z$}   \\
\tableline
      & $t$ = 9.0 & 9.5 & 10.0 & 10.5 & 11.0 & 11.5 & 12.0 & 12.5 & 13.0 & 13.5  & 14.0  \\
\tableline
--2.5 & 0.614 & 0.623 & 0.631 & 0.652 & 0.668 & 0.696 & 0.723 & 0.745 & 0.766 & 0.779 & 0.792 \\
--2.4 & 0.634 & 0.641 & 0.647 & 0.664 & 0.678 & 0.705 & 0.731 & 0.753 & 0.774 & 0.789 & 0.803 \\
--2.3 & 0.654 & 0.659 & 0.664 & 0.677 & 0.689 & 0.714 & 0.738 & 0.761 & 0.783 & 0.799 & 0.815 \\
--2.2 & 0.675 & 0.679 & 0.682 & 0.691 & 0.700 & 0.723 & 0.746 & 0.769 & 0.793 & 0.810 & 0.828 \\
--2.1 & 0.696 & 0.699 & 0.702 & 0.706 & 0.712 & 0.733 & 0.755 & 0.779 & 0.802 & 0.821 & 0.841 \\
--2.0 & 0.718 & 0.720 & 0.722 & 0.722 & 0.725 & 0.745 & 0.764 & 0.789 & 0.813 & 0.833 & 0.854 \\
--1.9 & 0.741 & 0.742 & 0.744 & 0.740 & 0.740 & 0.757 & 0.774 & 0.799 & 0.824 & 0.846 & 0.868 \\
--1.8 & 0.765 & 0.766 & 0.767 & 0.760 & 0.756 & 0.771 & 0.785 & 0.811 & 0.837 & 0.860 & 0.883 \\
--1.7 & 0.790 & 0.791 & 0.793 & 0.783 & 0.775 & 0.786 & 0.798 & 0.824 & 0.850 & 0.874 & 0.899 \\
--1.6 & 0.816 & 0.818 & 0.821 & 0.808 & 0.797 & 0.804 & 0.811 & 0.839 & 0.865 & 0.891 & 0.916 \\
--1.5 & 0.843 & 0.847 & 0.851 & 0.838 & 0.824 & 0.825 & 0.827 & 0.856 & 0.882 & 0.909 & 0.934 \\
--1.4 & 0.873 & 0.879 & 0.884 & 0.873 & 0.857 & 0.851 & 0.847 & 0.875 & 0.901 & 0.928 & 0.953 \\
--1.3 & 0.904 & 0.912 & 0.921 & 0.913 & 0.900 & 0.885 & 0.872 & 0.899 & 0.924 & 0.950 & 0.974 \\
--1.2 & 0.938 & 0.949 & 0.960 & 0.959 & 0.954 & 0.932 & 0.908 & 0.930 & 0.951 & 0.976 & 0.998 \\
--1.1 & 0.975 & 0.989 & 1.003 & 1.010 & 1.017 & 0.996 & 0.963 & 0.973 & 0.984 & 1.006 & 1.025 \\
--1.0 & 1.016 & 1.032 & 1.049 & 1.062 & 1.079 & 1.071 & 1.057 & 1.036 & 1.029 & 1.042 & 1.055 \\
--0.9 & 1.060 & 1.079 & 1.098 & 1.115 & 1.134 & 1.138 & 1.144 & 1.118 & 1.092 & 1.088 & 1.092 \\
--0.8 & 1.107 & 1.128 & 1.147 & 1.166 & 1.185 & 1.194 & 1.207 & 1.194 & 1.169 & 1.147 & 1.136 \\
--0.7 & 1.158 & 1.178 & 1.197 & 1.215 & 1.232 & 1.243 & 1.257 & 1.255 & 1.245 & 1.220 & 1.195 \\
--0.6 & 1.211 & 1.229 & 1.247 & 1.262 & 1.276 & 1.289 & 1.302 & 1.308 & 1.311 & 1.297 & 1.271 \\
--0.5 & 1.263 & 1.279 & 1.294 & 1.307 & 1.318 & 1.332 & 1.344 & 1.356 & 1.367 & 1.365 & 1.356 \\
--0.4 & 1.313 & 1.327 & 1.340 & 1.351 & 1.360 & 1.374 & 1.385 & 1.400 & 1.417 & 1.424 & 1.431 \\
--0.3 & 1.360 & 1.372 & 1.384 & 1.393 & 1.402 & 1.415 & 1.426 & 1.443 & 1.462 & 1.475 & 1.489 \\
--0.2 & 1.403 & 1.414 & 1.425 & 1.435 & 1.444 & 1.457 & 1.468 & 1.485 & 1.504 & 1.521 & 1.538 \\
--0.1 & 1.444 & 1.454 & 1.465 & 1.475 & 1.486 & 1.499 & 1.510 & 1.527 & 1.544 & 1.562 & 1.581 \\
 0.0 & 1.481 & 1.491 & 1.502 & 1.514 & 1.527 & 1.539 & 1.552 & 1.567 & 1.582 & 1.601 & 1.619 \\
 0.1 & 1.516 & 1.527 & 1.537 & 1.551 & 1.565 & 1.578 & 1.593 & 1.606 & 1.619 & 1.637 & 1.655 \\
 0.2 & 1.549 & 1.560 & 1.571 & 1.585 & 1.600 & 1.614 & 1.629 & 1.642 & 1.654 & 1.671 & 1.689 \\
 0.3 & 1.580 & 1.592 & 1.603 & 1.617 & 1.631 & 1.645 & 1.661 & 1.674 & 1.687 & 1.705 & 1.722 \\
 0.4 & 1.610 & 1.622 & 1.633 & 1.646 & 1.659 & 1.673 & 1.688 & 1.704 & 1.719 & 1.737 & 1.755 \\
 0.5 & 1.639 & 1.651 & 1.662 & 1.673 & 1.683 & 1.697 & 1.711 & 1.729 & 1.749 & 1.769 & 1.790 \\
\tableline
\end{tabular}
\tablecomments{The spectro-photometric model data (various colors and absorption indices) of the entire parameter space are available at http://web.yonsei.ac.kr/cosmic/data/YEPS.htm.}

\end{center}
\end{table}

\clearpage
\begin{table}
\footnotesize
\begin{center}
\caption{The median photometric errors in $g-z$ and $u-g$ of five magnitude bins for the M87 and M84 GCs \label{tbl-3}}
\begin{tabular}{c c c c}
\hline
\hline
Mag bins & Number of GCs & {\,\,\,\,\,\,\,\,\,\,$g-z$ error} & {\,\,\,\,\,\,\,\,\,\,$u-g$ error}  \\
\hline
\multicolumn{4}{c}{M87 ($u$ = WFPC2 F336W)}  \\
\hline
            $u_0$ $\le$ 23.0  &  20  &  {\,\,\,\,\,\,\,\,\,\,0.019}  &   {\,\,\,\,\,\,\,\,\,\,0.024} \\
23.0 $<$ $u_0$ $\le$ 24.0  &  87  &     {\,\,\,\,\,\,\,\,\,\,0.022}  &   {\,\,\,\,\,\,\,\,\,\,0.028} \\
24.0 $<$ $u_0$ $\le$ 25.0  &  172  &     {\,\,\,\,\,\,\,\,\,\,0.030}  & {\,\,\,\,\,\,\,\,\,\,0.054} \\
25.0 $<$ $u_0$ $\le$ 26.0  &  247  &     {\,\,\,\,\,\,\,\,\,\,0.043}  & {\,\,\,\,\,\,\,\,\,\,0.110} \\
            $u_0$ $>$  26.0                  & 65     &     {\,\,\,\,\,\,\,\,\,\,0.058}  &  {\,\,\,\,\,\,\,\,\,\,0.175}    \\
\hline
Entire Sample & 591   &   {\,\,\,\,\,\,\,\,\,\,0.035}   &  {\,\,\,\,\,\,\,\,\,\,0.078}  \\
\hline
\hline
\multicolumn{4}{c}{M84 ($u$ = WFC3 F336W)}  \\
\hline
            $u_0$ $\le$ 23.5     &  9      &     {\,\,\,\,\,\,\,\,\,\,0.019}  &   {\,\,\,\,\,\,\,\,\,\,0.021} \\
23.5 $<$ $u_0$ $\le$ 24.5  &  65    &     {\,\,\,\,\,\,\,\,\,\,0.025}  &  {\,\,\,\,\,\,\,\,\,\,0.032} \\
24.5 $<$ $u_0$ $\le$ 25.5  &  107  &     {\,\,\,\,\,\,\,\,\,\,0.035}  &  {\,\,\,\,\,\,\,\,\,\,0.059} \\
25.5 $<$ $u_0$ $\le$ 26.5  &  109  &     {\,\,\,\,\,\,\,\,\,\,0.057}  &  {\,\,\,\,\,\,\,\,\,\,0.111} \\
            $u_0$ $>$ 26.5       & 16     &     {\,\,\,\,\,\,\,\,\,\,0.082}  & {\,\,\,\,\,\,\,\,\,\,0.186}    \\
\hline
Entire Sample & 306   &   {\,\,\,\,\,\,\,\,\,\,0.039}   &  {\,\,\,\,\,\,\,\,\,\,0.068}  \\
\hline
\end{tabular}
\end{center}
\end{table}

\clearpage
\begin{table}
\scriptsize
\begin{center}
\caption{The $g-z$ color histograms (Figures 3$a$ \& 4$a$) of GCs in ACS VCS galaxies binned by host galaxy $B$-band magnitudes ($M_B$).\label{tbl-4}}
\vspace{0.5cm}
\begin{tabular}{cccccccc}
\tableline
\tableline
$g-z$ & \multicolumn{7}{c}{N} \\
\tableline
       & $-$22\,$\le$$M_B$$<$$-$21 & $-$21 $\sim$ $-$20 & $-$20 $\sim$ $-$19 & $-$19 $\sim$ $-$18 & $-$18 $\sim$ $-$17 & $-$17 $\sim$ $-$16 & $-$16 $\sim$ $-$15 \\
\tableline
0.35 		& {\,\,\,\,\,\,0 $\pm$ 0}       	& { \,2 $\pm$ 1}        	& { \,1 $\pm$ 1}       	& { \,1 $\pm$ 1}        	& { \,0 $\pm$ 0}		& { \,0 $\pm$ 0} 	& { \,0 $\pm$ 0} 	\\
0.45 		& {\,\,\,\,\,\,0 $\pm$ 1}       	& { \,3 $\pm$ 2}        	& { \,1 $\pm$ 1}       	& { \,0 $\pm$ 1}        	& { \,2 $\pm$ 1}		& { \,3 $\pm$ 2} 	& { \,0 $\pm$ 1} 	\\
0.55 		& {\,\,\,\,\,\,2 $\pm$ 3}       	& {\,$-$2 $\pm$ 3\,\,}  	& {\,$-$3 $\pm$ 4\,\,}  	& {\,$-$7 $\pm$ 4\,\,}  	& {\,$-$3 $\pm$ 4\,\,}   	& {\,$-$7 $\pm$ 5\,\,} 	& {\,$-$6 $\pm$ 4\,\,}	\\
0.65 		& {\,\,\,21 $\pm$ 5}     		& 18 $\pm$ 5      	& 16 $\pm$ 5     	& 19 $\pm$ 6     	& 12 $\pm$ 5    		& 22 $\pm$ 7  		& 12 $\pm$ 5 		\\
0.75 		& {\,\,\,85 $\pm$ 9}     		& 126 $\pm$ 11  	& 81 $\pm$ 9     	& 76 $\pm$ 9     	& 52 $\pm$ 8     	& 57 $\pm$ 9   		& 38 $\pm$ 7 		\\
0.85 		& {\,\,\,290 $\pm$ 17} 		& 389 $\pm$ 20 	& 294 $\pm$ 17 	& 200 $\pm$ 15 	& 171 $\pm$ 14 	& 159 $\pm$ 14   	& { \,99 $\pm$ 11} 	\\
0.95 		& {\,\,\,390 $\pm$ 20}		& 441 $\pm$ 21	& 295 $\pm$ 18 	& 188 $\pm$ 15 	& 152 $\pm$ 13 	& 120 $\pm$ 13	& { \,56 $\pm$ 10} 	\\
1.05 		& {\,\,\,328 $\pm$ 18} 		& 388 $\pm$ 20  	& 195 $\pm$ 14 	& 147 $\pm$ 13  	& 109 $\pm$ 11  	& { \,74 $\pm$ 10}  	& 30 $\pm$ 7		\\
1.15 		& {\,\,\,276 $\pm$ 16} 		& 397 $\pm$ 20 	& 217 $\pm$ 15 	& 133 $\pm$ 12  	& 78 $\pm$ 9  		&47 $\pm$ 8  	 	& 21 $\pm$ 6		\\
1.25 		& {\,\,\,323 $\pm$ 18} 		& 396 $\pm$ 20  	& 201 $\pm$ 14  	& 140 $\pm$ 13  	& 51 $\pm$ 8  		&33 $\pm$ 8  		& { \,2 $\pm$ 5}		\\
1.35 		& {\,\,\,416 $\pm$ 20} 		& 413 $\pm$ 20 	& 186 $\pm$ 14  	& 111 $\pm$ 11  	& 21 $\pm$ 5  		& 13 $\pm$ 5  		& { \,4 $\pm$ 4}		\\
1.45 		& {\,\,\,494 $\pm$ 22} 		& 322 $\pm$ 18 	& 137 $\pm$ 12  	& 70 $\pm$ 9  		& { \,4 $\pm$ 5}  	& {\,$-$1 $\pm$ 5\,\,} 	& {\,$-$7 $\pm$ 4\,\,}	\\
1.55 		& {\,\,\,373 $\pm$ 19} 		& 162 $\pm$ 13 	& 61 $\pm$ 8  		& 18 $\pm$ 5  		& { \,5 $\pm$ 3}  	& { \,0 $\pm$ 4}  	& { \,0 $\pm$ 3}		\\
1.65 		& {\,\,\,156 $\pm$ 12}		& 55 $\pm$ 8 	 	& 20 $\pm$ 5  		& { \,4 $\pm$ 4}  	& { \,1 $\pm$ 3}  	& { \,0 $\pm$ 4}  	& { \,0 $\pm$ 3}		\\
1.75 		& {\,\,\,56 $\pm$ 7} 		& 16 $\pm$ 5  		& { \,0 $\pm$ 3}  	& { \,1 $\pm$ 4} 	& {\,$-$2 $\pm$ 2\,\,} 	& { \,0 $\pm$ 3}  	& {\,$-$4 $\pm$ 3\,\,}	\\
1.85 		& {\,\,$-$2 $\pm$ 1} 		& {\,$-$3 $\pm$ 1\,\,}  	& {\,$-$5 $\pm$ 2\,\,}  	& {\,$-$7 $\pm$ 2\,\,}  	& {\,$-$4 $\pm$ 2\,\,} 	& {\,$-$6 $\pm$ 2\,\,} 	& {\,$-$6 $\pm$ 2\,\,}	\\
\tableline
\end{tabular}
\end{center}
\end{table}

\clearpage
\begin{table}
\scriptsize
\begin{center}
\caption{The inferred MDFs  (Figures 3$d$ \& 4$d$, and Figures 5 \& 7) of GCs in ACS VCS galaxies binned by host galaxy $B$-band magnitude ($M_B$).\label{tbl-5}}
\vspace{0.5cm}
\begin{tabular}{rccccccc}
\tableline
\tableline
{[Fe/H]}  &  \multicolumn{7}{c}{N/N$_{MAX}$} \\
\tableline
       & $-$22\,$\le$$M_B$$<$$-$21 & $-$21 $\sim$ $-$20 & $-$20 $\sim$ $-$19 & $-$19 $\sim$ $-$18 & $-$18 $\sim$ $-$17 & $-$17 $\sim$ $-$16 & $-$16 $\sim$ $-$15 \\
\tableline
$-$2.5 & 0.051 & 0.060 & 0.079 & 0.084 & 0.094 & 0.140 & 0.121 \\
$-$2.4 & 0.059 & 0.070 & 0.094 & 0.096 & 0.109 & 0.159 & 0.136 \\
$-$2.3 & 0.069 & 0.082 & 0.109 & 0.113 & 0.129 & 0.184 & 0.158 \\
$-$2.2 & 0.081 & 0.095 & 0.127 & 0.130 & 0.153 & 0.214 & 0.183 \\
$-$2.1 & 0.094 & 0.110 & 0.148 & 0.150 & 0.180 & 0.247 & 0.213 \\
$-$2.0 & 0.112 & 0.127 & 0.174 & 0.173 & 0.214 & 0.288 & 0.250 \\
$-$1.9 & 0.131 & 0.147 & 0.201 & 0.198 & 0.253 & 0.335 & 0.293 \\
$-$1.8 & 0.151 & 0.169 & 0.233 & 0.225 & 0.297 & 0.382 & 0.345 \\
$-$1.7 & 0.175 & 0.195 & 0.266 & 0.255 & 0.350 & 0.439 & 0.412 \\
$-$1.6 & 0.202 & 0.224 & 0.303 & 0.288 & 0.413 & 0.501 & 0.494 \\
$-$1.5 & 0.231 & 0.255 & 0.344 & 0.325 & 0.482 & 0.573 & 0.592 \\
$-$1.4 & 0.264 & 0.290 & 0.389 & 0.368 & 0.560 & 0.648 & 0.696 \\
$-$1.3 & 0.300 & 0.330 & 0.436 & 0.417 & 0.643 & 0.734 & 0.799 \\
$-$1.2 & 0.339 & 0.379 & 0.484 & 0.477 & 0.729 & 0.837 & 0.890 \\
$-$1.1 & 0.378 & 0.440 & 0.534 & 0.552 & 0.834 & 0.939 & 0.962 \\
$-$1.0 & 0.424 & 0.518 & 0.608 & 0.636 & 0.954 & 0.999 & 1.000 \\
$-$0.9 & 0.484 & 0.637 & 0.750 & 0.760 & 0.997 & 0.970 & 0.822 \\
$-$0.8 & 0.589 & 0.840 & 0.954 & 0.954 & 0.900 & 0.781 & 0.331 \\
$-$0.7 & 0.764 & 0.994 & 0.983 & 0.948 & 0.581 & 0.470 & 0.127 \\
$-$0.6 & 0.940 & 0.945 & 0.858 & 0.742 & 0.281 & 0.249 & 0.068 \\
$-$0.5 & 1.000 & 0.808 & 0.674 & 0.543 & 0.164 & 0.118 & 0.024 \\
$-$0.4 & 0.948 & 0.615 & 0.498 & 0.369 & 0.094 & 0.046 & 0.003 \\
$-$0.3 & 0.802 & 0.424 & 0.332 & 0.236 & 0.048 & 0.009 & 0.000 \\
$-$0.2 & 0.601 & 0.278 & 0.214 & 0.142 & 0.033 & 0.000 & 0.000 \\
$-$0.1 & 0.420 & 0.176 & 0.141 & 0.080 & 0.037 & 0.000 & 0.001 \\
0.0 & 0.280 & 0.111 & 0.092 & 0.044 & 0.036 & 0.007 & 0.015 \\
0.1 & 0.184 & 0.071 & 0.055 & 0.020 & 0.017 & 0.005 & 0.007 \\
0.2 & 0.121 & 0.044 & 0.028 & 0.007 & 0.002 & 0.000 & 0.000 \\
0.3 & 0.078 & 0.028 & 0.011 & 0.002 & 0.000 & 0.000 & 0.000 \\
0.4 & 0.047 & 0.017 & 0.002 & 0.001 & 0.000 & 0.000 & 0.000 \\
0.5 & 0.025 & 0.010 & 0.000 & 0.000 & 0.000 & 0.000 & 0.000 \\
\tableline
\end{tabular}
\end{center}
\end{table}

\end{document}